\documentclass[]{aa}
\usepackage[dvips]{graphicx}
\usepackage{amssymb}
\newcommand{\mcol}[3]{\multicolumn{#1}{#2}{#3}}

\newcommand{\Teff}{\mbox{$T_{\mbox{\rm \tiny eff}}$}}

\begin{document}

\title{The Modelling of Intermediate Age Stellar Populations} 

\subtitle{II. Average Spectra for Upper AGB Stars, and their Use}

\author{A.\,Lan\c{c}on and M.\,Mouhcine\inst{1}}

\offprints{A. Lan\c{c}on}

\institute{Observatoire Astronomique, Universit\'e L.\,Pasteur \& CNRS: 
UMR 7550, 11 rue de l'Universit\'e, F-67000 Strasbourg}

\date{Received 19 June 2001 / Accepted 15 April 2002}

\authorrunning{A. Lan\c{c}on \& M. Mouhcine}
\titlerunning{Average Spectra for Upper AGB Stars}

\abstract{
The upper Asymptotic Giant Branch (AGB) is populated with oxygen rich and 
carbon rich Long Period Variables (LPVs). These stars are essential contributors
to the near-IR light of intermediate age stellar populations. 
Individual observed spectra of LPVs are so diverse that they cannot be used 
directly in the synthesis of galaxy spectra. In this paper, the library of 
individual spectra of Lan\c{c}on \& Wood (2000) is used to construct averages
that can be incorporated conveniently in population synthesis
work. The connection between such spectra and stellar evolution tracks
is discussed.
In order to select a sorting criterion and to define averaging bins for the 
LPV spectra, correlations between their spectrophotometric properties
are reexamined. While optical properties and
broad baseline colours such as (I-K) are well correlated, 
a large dispersion is observed when these indices are plotted
against near-IR ones. This is partly due to the intrinsic width
of the upper AGB, which is illustrated by locating each of the multiple
observations of individual LPVs on the HR diagram. It is argued that
broad baseline colour-temperatures are the most sensible sorting
criteria. The properties of the resulting sequence of average
spectra indeed vary regularly.
We further address: (i) the bolometric corrections and temperature
scales needed to associate a spectrum with a given point on
a theoretical stellar evolution track (or isochrone), (ii) the
simplifying assumptions that will be implicitely made when using
the average spectra, (iii) potential biases in the sample
of Lan\c{c}on \& Wood and their effects, (iv) the small contribution
of LPVs to the interstellar hydrogen emission lines in galaxies. It
is emphasized that an a posteriori calibration of the effective 
temperature scale remains necessary, until consistent models for the 
evolution, the pulsation and the spectral appearance of LPVs become available. 
We suggest a recipe for the use of the average spectra at various metallicities.
\keywords{Stars: AGB and post-AGB -- Stars: late-type --
Stars: variable: general -- Hertzsprung-Russell diagram --
Infrared: stars -- Galaxies: stellar content} 
}

\maketitle

\vspace*{-0.4cm}
\section{Introduction}
\label{intro.sec}

The prediction of the near-infrared (near-IR) spectra of star clusters and
galaxies is subordonate to the existence of stellar spectral 
libraries with a complete coverage of the evolved stages of stellar
evolution. Stellar libraries are also necessary in the construction of
near-IR colour-magnitude or two-colour diagrams, 
which may include narrow-band 
filter indices or absorption line measurements. A shortcoming of all
libraries previously used in population synthesis work 
(e.g. Kleinmann \& Hall \cite{KH86}, Terndrup et al. \cite{TFW91},
Lan\c{c}on \& Rocca-Volmerange \cite{LRV92}, Pickles \cite{Pick98}) 
is the poor coverage of the upper Asymptotic Giant Branch (AGB).
The upper AGB hosts oxygen-rich and carbon-rich Long Period Variables 
(LPVs)\footnote{
In agreement with common usage,
LPVs include both semi-regular and Mira-type variables; Miras are
LPVs with large (optical) amplitudes, i.e. $\delta$V$>$2.5\,magnitudes
(Kholopov et al. \cite{GCVS85}, Lloyd Evans \cite{Llo83}, 
Hughes \& Wood \cite{HW90}).},
whose specific spectral features carry the potential of revealing
the presence of intermediate age populations (Lan\c{c}on et al. \cite{LMFS99}).
Indeed, upper AGB stars contribute of the order of 50\,\% of the K band
light over a range of stellar population ages between $\sim 10^8$ and 
$\sim 10^9$\,years (Persson et al. \cite{PACFM83}, 
Renzini \& Buzzoni \cite{RB86}, Frogel et al. \cite{FMB90},
Ferraro et al. \cite{FFTetal95}, Bressan et al. \cite{BGS98},
Girardi \& Bertelli \cite{GB98}, Lan\c{c}on \cite{Lan98}, 
Maraston \cite{Mara98}, Mouhcine \& Lan\c{c}on \cite{ML02_models}).

Recently, Lan\c{c}on \& Wood (\cite{LW00}; hereafter LW2000) 
published a library of stellar spectra of luminous cool stars
that includes a large sample of instantaneous observations of LPVs.
The data span the wavelength range where most of the light of 
cool stars is emitted (0.5 to 2.5\,$\mu$m) with a spectral
resolving power of $\sim 1100$ longward of 1\,$\mu$m, and,
with about 100 spectra, provide the most complete data set available
to date.
 
The range of spectrophotometric properties observed in LPVs is large
(LW2000, Alvarez et al. \cite{ALPW00}). 
Even the classification of the empirical spectra has many caveats. 
Sorting algorithms based on optical or on near-IR criteria, 
on broad band colours or on spectroscopic 
properties, on mean or on instantaneous data, often give 
different results. The most conspicuous features, such as the 
H$_2$O vapour absorption bands around 1.4\,$\mu$m and 1.9\,$\mu$m
(in oxygen rich LPVs), depend on details of the atmospheric structure
that change from one pulsation cycle to the next.
Such changes will be smoothed out in the integrated light of
(large) stellar populations. 

The purpose of the present paper is to provide a library of {\em mean}
stellar spectra of luminous cool variable stars, 
{\em suitable for direct use in combination with a population synthesis model}. The library complements more readily available data for
static red giants or supergiants. Both oxygen rich
and carbon rich AGB stars are considered. We attempt to thoroughly
discuss the impact of technical/physical choices that have to
be made at various steps in the preparation and the use of 
the averaged spectra.
\medskip

The construction of a library of mean spectra
requires the preliminary construction of suitable averaging bins. 
In principle, one would like to compute
energy weighted mean spectra for individual stars, based on 
spectra taken at various phases over several pulsation cycles and
on the light curve. However, this requires an enormous amount of 
spectroscopic and photometric data. Even the LW2000 data are not sufficient.
This paper suggests that colour temperature is the best single
parameter to order LPV spectra.
The sample of useful input spectra is briefly described in 
Sect.\,\ref{sample.sec}, where the main selection criteria 
(and thus potential selection biases) are also recalled. 
Sequences of average spectra are presented 
in Sect.\,\ref{bin_choice.sec}. The statistical properties of the 
data that justify our selected sorting criteria are also
presented in that section.
\medskip

The discussion of the uncertainties in population synthesis
predictions inherent to the chosen averaging
procedure must be based on the 
understanding of two processes that tend to disperse the 
stars of even a coeval population over quite a broad area of the 
Hertzsprung-Russell (HR) diagram: the TP-AGB thermal pulse cycles and
the LPV pulsation cycles. 
The theoretical stellar evolution tracks used 
in population synthesis calculations
provide the time evolution of the position of a star in the HR diagram.
The thermal pulses are now commonly accounted for in the tracks, 
but the effects of pulsation, on timescales of hundreds or thousands 
of days, are not included. 
One may think of these tracks as providing the evolution of the 
static parent stars of LPVs. 
Unfortunately, building non-linear pulsation models and theoretical spectra 
for LPVs is extremely complex, 
and current models cannot yet provide the fundamental
parameters of the parent star on the basis of one or several 
instantaneous empirical spectra or colours.

How, then, should one connect the averaged spectra of the new 
library with locations along the evolutionary tracks?
Practical aspects of this question are
addressed in Sect.\,\ref{useit.sec}. The subsequent discussion
addresses fundamental difficulties. In particular,
it emphasizes the need for a posteriori calibration of the
relation between colour temperatures and effective temperatures
(i.e. the temperature scale of the spectra). 
Potential selection biases in the LW2000 data, and their effect,
are also discussed in Sect.\,\ref{disc.sec}.
Recipes for extensions of the population synthesis calculations
to other metallicities than quasi-solar are provided
in Sect.\,\ref{metal.sec}. A concluding summary is given
in Sect.\,\ref{concl.sec}.
\medskip

The averaged spectra are available in digital form
through CDS\footnote{Centre
de Donn\'ees Astrophysiques de Strasbourg,
{\tt http://cdsarc.u-strasbg.fr/CDS.html}, VizieR service}.

\section{The sample}
\label{sample.sec}

The library of LW2000 contains more than 100 merged optical/near-IR
spectra. Of these, about one quarter
belong to giants of the first giant branch, red supergiants, 
Galactic Bulge LPVs or luminous red stars of the Large and Small
Magellanic Clouds. The remaining spectra are those of pulsating 
giants in the field of the Milky Way. These are the data we
focus on.

The sample of LW2000 was selected to provide a relatively
uniform coverage of the period--amplitude plane of LPVs (their Fig.\,1). 
The O-rich LPVs with V band amplitudes ($\delta$V) smaller than 0.7\,magnitudes
in the sample show no obvious difference with static giants.
We exclude these objects from the subsample used here.
The resulting set contains 63 O-rich spectra of LPVs with 
$\delta$V\,$\geq$\,0.8\,mag.

The sample of C-rich stars contains only 6 objects, of which one
is an S/C star (3 of 21 spectra) and one is a far-IR
source (R\,Lep\,=\,IRAS 04573-1452; 3 spectra). 
Although their amplitudes are larger than 1.5\,mag 
in V, the spectra show relatively little variation with phase. 
The largest amplitude amounts $\delta$V$\simeq$6\,mag
for the far-IR source.
The next largest amplitude is $\delta$V$\simeq$3.7. 
The spectrophotometric properties
of the C stars in the sample are much less dispersed than 
those of the O-rich LPVs.

\section{Average spectra for LPVs}
\label{bin_choice.sec}

In principle, the ideal spectral
library for population synthesis purposes should consist of a large
collection of light-curve weighted spectra of individual stars observed
photometrically and spectroscopically over several pulsation cycles.
There are two main reasons why this is not practical. First, the
amount of data required is enormous. In the sample of LW2000, only
a few light curves have been sampled at more than 4 points. The 
actual light curves and phases are generally not accurately known. No
more complete empirical spectral library for LPVs is available in
the literature. Second, uncertainties in  current models make it 
impossible to determine the physical nature of one particular star 
(initial mass, current evolutionary status), even when
many spectra are available for it. Attempts to follow the ``ideal" route 
fail at a very basic level: the mean spectra of individual stars 
cannot be sorted into a sequence that behaves regularly enough for 
practical use. 

Along the TP-AGB, the effective temperature globally decreases while the 
luminosity increases. In this section, we show that the use of effective 
temperature indicators as basic sorting criteria produces a 
sequence of O-rich spectra with regularly evolving properties. 
The tightness of this empirical sequence 
tells us that surface temperature is indeed a good parameter  
for the first order classification of the spectra. Other 
sorting parameters are discussed but are found to be less 
appropriate. Using the suggested temperature sequence is, in our opinion, 
the most sensible way to include O-rich LPVs in evolutionary
synthesis models for galaxies.

In the case of C-rich stars, both a temperature sequence and 
a sequence of C/O abundance ratio are considered.
 
\subsection{Oxygen-rich LPVs}

\subsubsection{Temperature sequence}
\label{O_sequence.sec}

The spectra of oxygen-rich LPVs selected in Sect.\,\ref{sample.sec}
cover a range of 8\,magnitudes in (V-K) and 4\,magnitudes
in (V-I) or (I-K)\footnote{Unless otherwise stated, broad band colours
are measured on the reduced spectra using the transmission
curves of Bessell (\cite{Bes90}) for V, R and I (Cousins system), and
of Bessell \& Brett (\cite{BB88}) for J, H and K (Johnson-Glass system).
Zero colours are adopted for Vega, based on the Vega model spectrum
also used by Bessell \& Brett (we thank M.\,Bessell for providing
the latter).}. 
These long baseline colours are expected to correlate
with effective temperature (Bessell et al. \cite{BBSW89b},
\cite{BCP98}), although the absolute  calibration of the relation is 
poorly known for variable stars, and is metallicity-dependent 
(cf. Sect.\,\ref{metal.sec}). 

\begin{figure}[!htb]
\includegraphics[clip=,width=0.5\textwidth]{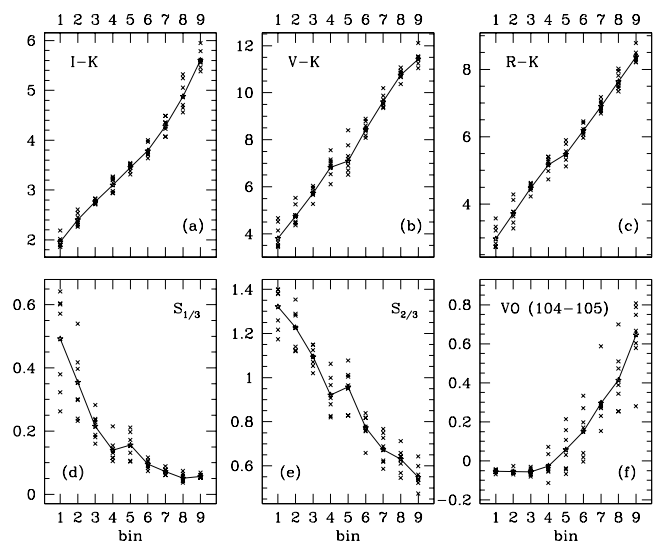}
\caption[]{Temperature sensitive indices as a function of the
bin number based on (I-K). The indices of the O-rich LPVs of LW2000 
are shown as crosses; the solid lines connect those of the averaged spectra.
All colours are in magnitudes, with a value of 0 for Vega. In {\bf (b)},
V is measured through a standard filter truncated at 5100\,\AA
(cf. Alvarez et al. \cite{ALPW00}).} 
\label{ImK7_disp_Tcols.fig}
\end{figure}

Figure\,\ref{ImK7_disp_Tcols.fig} confirms that tight correlations
exist between various commonly used photometric temperature indicators.
The small dispersion in panels (b) and (c), which relate (V-K)
and (R-K) to (I-K), shows that the three long baseline broad band colours 
contain equivalent information. In the following, 
(I-K) is adopted as the primary temperature index.
The choice between (I-K) and (R-K) is arbitrary, while
(V-K) was rejected because the passbands of standard V filters
extend significantly over the edge of the LW2000 data, requiring
extrapolation of the data or (as was done here) a
truncation of the filter passband. 

The VO index of panel\,(f) 
measures the depth of the VO A--X $(\Delta v = 0)$ 
band at 1.05\,$\mu$m, as done by Bessell et al. (\cite{BBSW89b}) and
Alvarez et al. (\cite{ALPW00}). The absence of the molecular 
band in the warmer LPV spectra restricts its use as a temperature
indicator.  S$_{1/3}$ and S$_{2/3}$ in panels (d) and (e)
are the optical narrow band filter indices defined by
Fluks et al. (\cite{FPTetal94})\footnote{Note that integrated flux ratios are
used in that definition, instead of the usual flux density ratios.}.
For static giants, they are good indicators of the spectral type,
although the molecular bands that determine S$_{1/3}$ saturate
for the coolest stars (spectral types later than M6). 
The relative dispersion in the figures involving narrow bands
is due to the sensitivity of individual molecular bands to
the structure of the outer atmosphere, which is
affected by details of pulsation. This sensitivity
has been discussed by Bessell et al. \cite{BBSW89b} on the basis
of Mira models and is found empirically in
measured variations of the bands relative to
each other or to the colours, from cycle to
cycle and with phase (Spinrad \& Wing \cite{SW69},
Alvarez \& Plez, \cite{AP98}, LW2000).
It strengthens the preference given to (I-K) as a temperature
indicator.

\begin{figure}[!hbt]
\includegraphics[clip=,angle=0,
        width=0.5\textwidth]{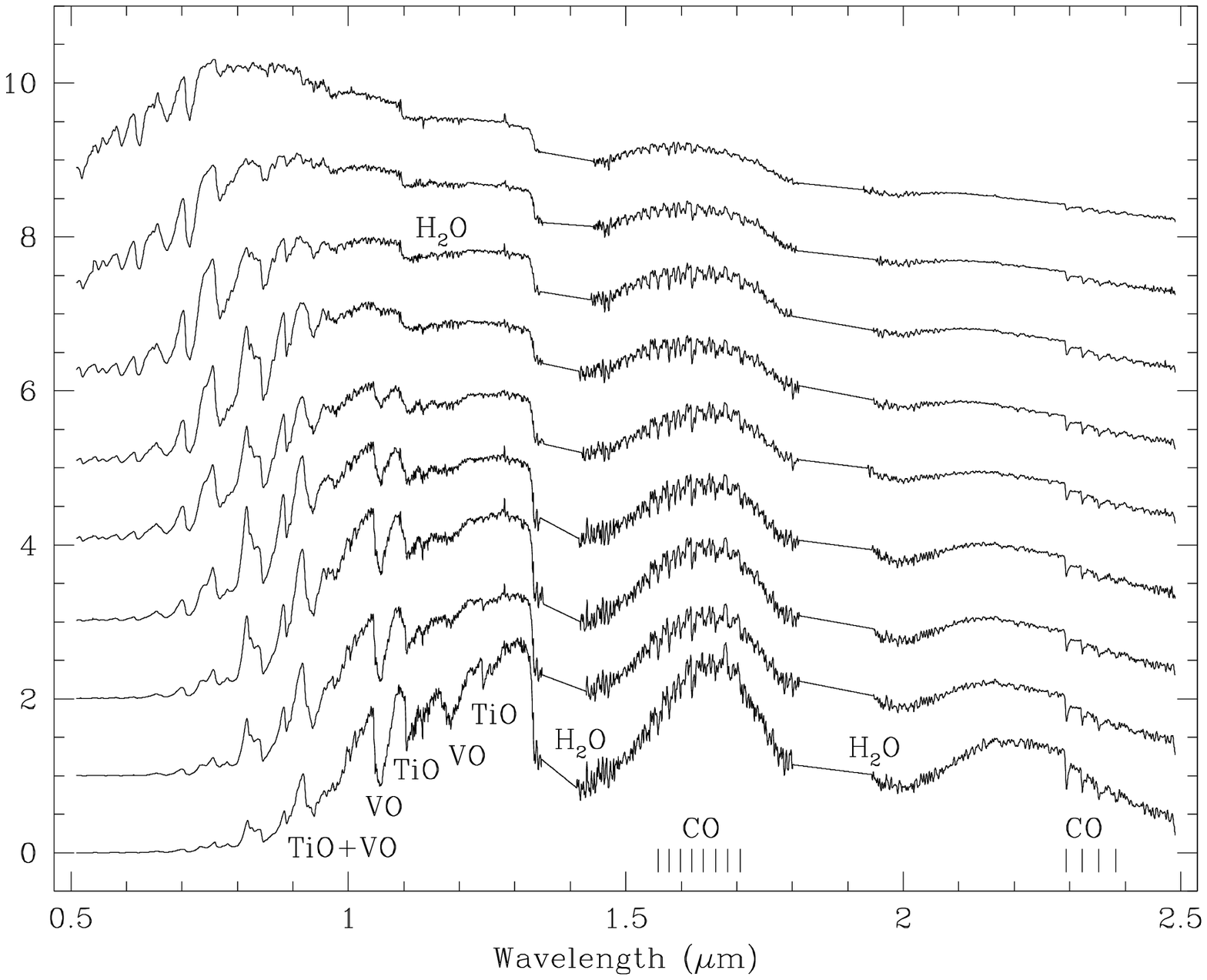}
\caption[]{The sequence of O-rich LPV spectra, obtained after sorting
the spectra according to (I-K). Linear flux density units are used
(F$_{\lambda}$). The spectra have been normalized to a common
integrated flux, and each is offset from the lower one by one
unit. Straight segments interpolate through the regions where the
telluric absorption was too strong for a satisfactory correction.}
\label{Miras.ImK7seq.fig}
\end{figure}

(I-K) was used to sort the 63 individual spectra into 9 bins 
of 7 spectra each, as listed in Table\,\ref{ImK7bins.tab} and 
illustrated in Fig.\,\ref{ImK7_disp_Tcols.fig}\,(a).
As a result of this choice, bin sizes in terms of (I-K) are not
constant. No binning choice guarantees constant bin sizes
in all colours, and choosing constant bin sizes in (I-K)  
rather than a constant number of spectra per bin 
would have given (I-K) more importance than it deserves 
compared to the other colours of Fig.\,\ref{ImK7_disp_Tcols.fig}.
Within each bin, the spectra were normalised to a common mean
flux over the whole wavelength range of the data, and were
then averaged without further weighting. The resulting sequence 
of average spectra is displayed in Fig.\ref{Miras.ImK7seq.fig}.
The maximum of the energy distribution is seen to move from
about 7500\,\AA\ for the hottest spectrum to about 1.3\,$\mu$m
for the coolest. Molecular bands grow more numerous and deeper
with decreasing colour temperature. The corresponding photometric
index measurements are located on the solid lines in 
Fig.\,\ref{ImK7_disp_Tcols.fig}. 

\begin{table*}[ht]
\small
\caption[]{Contents of the O-rich temperature bins}
\label{ImK7bins.tab}
\parbox[b]{0.5\textwidth}{
\begin{tabular}{crllccc} \hline
Bin & Name & Date & Type & Period & $\delta$V & (I-K) \\ 
(1) & (2) & (3) & (4) & (5) & (6) & (7) \\ \hline
1 & T\,Cen   & Jul.\,96   &  Sa   &        90.4   &       3.5  &   1.86 \\
1 & T\,Cen   & Jan.\,96   &  Sa   &        90.4   &       3.5  &   1.90 \\
1 & BD\,Hya  & Jul.\,96   &  Sa   &       117.4   &       1.8  &   1.90 \\
1 & S\,Car   & Jun.\,95   &  M    &       149.5   &       5.4  &   1.90 \\
1 & S\,Car   & Dec.\,95   &  M    &       149.5   &       5.4  &   1.98 \\
1 & T\,Cen   & Mar.\,96   &  Sa   &        90.4   &       3.5  &   2.01 \\
1 & RS\,Hor  & Mar.\,96   &  M    &       202.9   &       5.2  &   2.19 \\
2 & S\,Lib   & May\,96   &  M    &       192.9   &       5.5  &   2.26 \\
2 & S\,Lib   & Jun.\,95   &  M    &       192.9   &       5.5  &   2.29 \\
2 & S\,Car   & May\,96   &  M    &       149.5   &       5.4  &    2.32 \\
2 & T\,Cen   & May.\,96   &  Sa   &        90.4   &       3.5  &   2.36 \\
2 & U\,Crt   & Jan.\,96   &  M    &        169.   &        4.  &   2.45 \\
2 & BD\,Hya  & May\,96   &  Sa   &       117.4   &       1.8  &    2.53 \\
2 & BD\,Hya  & Jan.\,96   &  Sa   &       117.4   &       1.8  &    2.61 \\
3 & S\,Car   & Jan.\,96   &  M    &       149.5   &       5.4  &   2.71 \\
3 & R\,Phe   & May\,96   &  M    &       269.3   &       6.9  &   2.72 \\
3 & S\,Car   & Jul.\,96   &  M    &       149.5   &       5.4  &   2.73 \\
3 & S\,Phe   & Dec.\,95   &  Sb   &        141.   &        2.  &   2.75 \\
3 & UZ\,Hya  & Jan.\,96   &  M    &        261.   &       5.7  &   2.78 \\
3 & UZ\,Hya  & Mar.\,96   &  M    &        261.   &       5.7  &   2.82 \\
3 & R\,Phe   & Jul.\,96   &  M    &       269.3   &       6.9  &   2.83 \\
4 & U\,Crt   & May\,96   &  M    &        169.   &        4.  &   2.93 \\
4 & KV\,Car  & May\,96   &  Sb   &        150.   &       0.8  &   2.94 \\
4 & RS\,Hor  & Jan.\,96   &  M    &       202.9   &       5.2  &   2.98 \\
4 & RS\,Hya  & Jun.\,95   &  M    &       338.6   &       5.2  &   3.19 \\
4 & RS\,Hya  & May\,96   &  M    &       338.6   &       5.2  &   3.22 \\
4 & S\,Phe   & Jan.\,96   &  Sb   &        141.   &        2.  &   3.24 \\
4 & SY\,Vel  & Mar.\,96   &  Sb   &         63.   &       1.3  &    3.27 \\
5 & KV\,Car  & Jan.\,96   &  Sb   &        150.   &       0.8  &   3.31 \\
5 & S\,Car   & Mar.\,96   &  M    &       149.5   &       5.4  &   3.38 \\
5 & SY\,Vel  & Jan.\,96   &  Sb   &         63.   &       1.3  &   3.45 \\
\hline
\end{tabular}

\rule[0pt]{0pt}{1mm}
}
\parbox[b]{0.5\textwidth}{
\begin{tabular}{crllccc} \hline
Bin & Name & Date & Type & Period & $\delta$V & (I-K) \\
(1) & (2) & (3) & (4) & (5) & (6) & (7) \\ \hline
5 & S\,Phe   & Jul.\,96   &  Sb   &        141.   &        2.  &   3.49 \\
5 & S\,Phe   & May\,96   &  Sb   &        141.   &        2.  &   3.51 \\
5 & RY\,Cra  & Jun.\,95   &  M    &        195.   &       1.9  &   3.51 \\
5 & RS\,Hya  & Jul.\,96   &  M    &       338.6   &       5.2  &   3.54 \\
6 & RS\,Lib  & May\,96   &  M    &       217.6   &        6.  &   3.63 \\
6 & RS\,Hya  & Apr.\,96   &  M    &       338.6   &       5.2  &   3.70 \\
6 & RZ\,Car  & May.\,96   &  M    &       272.8   &       6.2  &   3.73 \\
6 & SV\,Tel  & Jun.\,95   &  M    &       225.5   &        3.  &   3.76 \\
6 & R\,Phe   & Dec.\,95   &  M    &       269.3   &       6.9  &   3.77 \\
6 & RS\,Hor  & Dec.\,95   &  M    &       202.9   &       5.2  &   3.97 \\
6 & R\,Cha   & Dec.\,95   &  M    &       334.6   &       6.7  &   4.00 \\
7 & RS\,Hya  & Mar.\,96   &  M    &       338.6   &       5.2  &   4.07 \\
7 & X\,Men   & Jul.\,96   &  M    &        380.   &       3.6  &   4.07 \\
7 & R\,Phe   & Jan.\,96   &  M    &       269.3   &       6.9  &   4.23 \\
7 & R\,Cha   & May\,96   &  M    &       334.6   &       6.7  &   4.34 \\
7 & WW\,Sco  & May\,96   &  M    &        431.   &       4.1  &   4.36 \\
7 & RZ\,Car  & Jun.\,95   &  M    &       272.8   &       6.2  &   4.49 \\
7 & RZ\,Car  & Jul.\,96   &  M    &       272.8   &       6.2  &   4.49 \\
8 & SV\,Lib  & Jul.\,96   &  M    &       402.7   &       1.3  &   4.56 \\
8 & SV\,Lib  & Jun.\,95   &  M    &       402.7   &       1.3  &   4.65 \\
8 & SV\,Lib  & May\,96   &  M    &       402.7   &       1.3  &   4.70 \\
8 & R\,Cha   & Jan.\,96   &  M    &       334.6   &       6.7  &   4.86 \\
8 & CM\,Car  & Jan.\,96   &  M    &        335.   &       2.5  &   5.05 \\
8 & RS\,Hya  & Jan.\,96   &  M    &       338.6   &       5.2  &   5.24 \\
8 & X\,Men   & Dec.\,95   &  M    &        380.   &       3.6  &   5.32 \\
9 & RZ\,Car  & Mar.\,96   &  M    &       272.8   &       6.2  &   5.38 \\
9 & X\,Men   & Mar.\,96   &  M    &        380.   &       3.6  &   5.46 \\
9 & X\,Men   & Jan.\,96   &  M    &        380.   &       3.6  &   5.55 \\
9 & RZ\,Car  & Jan.\,96   &  M    &       272.8   &       6.2  &   5.57 \\
9 & WW\,Sco  & Jun.\,95   &  M    &        431.   &       4.1  &   5.61 \\
9 & RS\,Lib  & Jun.\,95   &  M    &       217.6   &        6.  &   5.79 \\
9 & R\,Cha   & Mar.\,96   &  M    &       334.6   &       6.7  &   5.95 \\
\hline
\end{tabular}

\rule[0pt]{0pt}{1mm}
}
\normalsize
\end{table*}

\subsubsection{Near-IR properties}
\label{nir_ppties.sec}

\begin{figure}[!ht]
\includegraphics[clip=,width=0.5\textwidth]{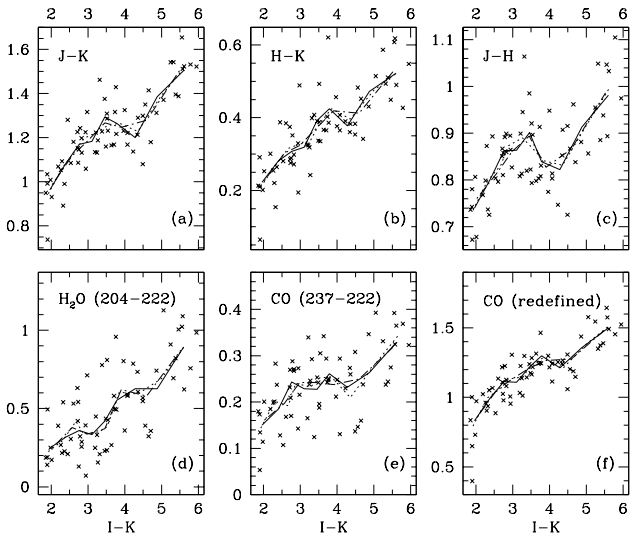}
\caption[]{Common near-IR indices as a function of (I-K), for O-rich LPVs.
Lines show the properties of averaged spectra, with the three bin definitions
given in the text. Plots {\bf (d)} and {\bf (e)} are flux density
ratios, measured through the bandpasses of the HST/\-NICMOS/\-Camera\,2
filters F204M, F222M, F237M, as labeled, and converted to magnitudes
(0 for Vega). The CO index in plot {\bf (f)} is colour-corrected (see text).}
\label{ImK_679_nir.fig}
\end{figure}

The spectral features of highest relevance to extragalactic studies
are located longward of 1$\,\mu$m, i.e.
at wavelengths where the relative contribution of the AGB stars to
the integrated light can be significant. The VO band of
Fig.\,\ref{ImK7_disp_Tcols.fig}\,(f) is one example.
The behaviour of the near-IR broad band colours (J-K), (H-K) and (J-H)
and of the CO and H$_2$O absorption bands is shown in
Fig.\,\ref{ImK_679_nir.fig}. All these indices are
based on the Hubble Space Telescope NICMOS filter passbands\footnote{
Many molecular index definitions are found in the 
literature. We recommend that users measure the index of
their choice on the actual spectra instead of attempting
a conversion from the plots in this section.}.
They are given in magnitudes, with the assumption that
all colours are 0 for Vega.
The solid lines refer to the properties of our prefered
sequence of averages, described in Sect\,\ref{O_sequence.sec}. The dashed
and dot-dashed lines are obtained respectively
with 7 bins of 9 spectra each, and with 10 bins of 6 or 7 spectra each.

Fig.\,\ref{ImK_679_nir.fig} shows that 
the near-IR broad band colours (J-K), (H-K) and (J-H) correlate with the
(I-K) colour-temperature, but with a large dispersion and a
relatively shallow slope. {\em If} (I-K) really is a good effective temperature
estimator for O-rich LPVs, then it is clear that the near-IR colours
are poor ones. Bins based on near-IR colours would mix spectra
with very diverse general energy distributions, optical spectra
and spectral types.

The observed decrease
in the near-IR colour indices between (I-K)=3.3 and (I-K)=4.5
(bins 5 to 7) is the result of the competition between the influences
of temperature and of molecular absorption. This
decrease is statistically significant in our sample,
in particular in (J-H): the size of the step is about 3 times the
estimated r.m.s. error on the mean colour (r.m.s. dispersion
of the data divided by $\sqrt{7}$, as we have 7 spectra per bin). 
However, it is unclear whether it would be seen in larger samples. 
In addition, the step is relatively small when compared to the
whole range of the data, and it is comparable to typical observational
uncertainties in near-IR photometric observations. Our summary
of these data is that the mean (J-K), (H-K) and (J-H)  of O-rich LPVs
are approximately constant between (I-K)$\simeq$3 and (I-K)$\simeq$5
(i.e. bins $\sim$3 to $\sim$8).

\begin{figure}[hbt]
\includegraphics[clip=,width=0.5\textwidth]{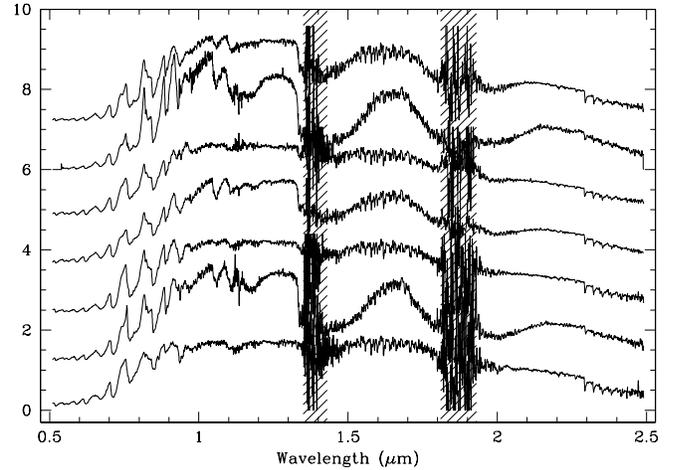}
\caption[]{The seven O-rich LPVs of bin number 5 (in the order
of Table\,\ref{ImK7bins.tab} from the bottom upwards). An illustration of
the variety of near-IR properties seen at a given (I-K).}
\label{Bin5.fig}
\end{figure}

Spectra with similar energy distributions display an
impressive variety of near-IR molecular absorption spectra,
as illustrated in Fig.\,\ref{Bin5.fig}. 
Again, this shows how sensitive the spectral signatures
of molecules such as H$_2$O, that form far in the outer atmosphere
(Matsuura et al. \cite{MYMO01}, models of Hofmann et al. \cite{HSW98}:
priv. communication by M. Scholz), 
are to the instantaneous pulsation-determined structure of the latter.
We will attempt to further interpret these differences in 
Sect.\,\ref{PdV_correl.sec}.
Differences in the  width and depth of the wings of the broad H$_2$O
absorption bands contribute significantly to the spread around
the mean near-IR broad band colours.
CO absorption in the H band plays a smaller role. Real
deviations in the pseudo-continuum between 1 and 1.3\,$\mu$m
occur (see Sect.\,3.5 of LW2000) and also contribute to the spread.

Along the sequence of averaged spectra, the water vapour absorption bands 
deepen and widen with decreasing temperature. The measurements
shown are based on the HST/\-NICMOS filters F204M and F222M:
$$ \rm{H}_2\rm{O} = 
	-2.5 \log [ \rm{F}_{\lambda} (F204M) / \rm{F}_{\lambda} (F222M) ]
	 - \rm{H}_2\rm{O} (\rm{Vega}) $$
The first is centered at 2.04\,$\mu$m, deep in one of the H$_2$O bands,
the second at 2.22\,$\mu$m, at a quasi-continuum point.
The lower envelope of the data points in Fig.\,\ref{ImK_679_nir.fig}\,(d) 
illustrates the mere effect of the change in colour-temperature for stars with 
shallow or inexistent H$_2$O bands. At a given (I-K),
star-to-star differences of up to 0.7\,magnitudes are seen in
the (204-222) index. Figure\,\ref{Bin5.fig} shows that similar spreads are
expected in other band wings of this molecule, thus 
affecting in particular the broad band photometry.
For instance, the star with the largest 
(H-K) (at (I-K)$\simeq$3.8) also has extreme water bands (second spectrum
from the top of Fig.\,\ref{Bin5.fig}).

The CO index in Fig.\,\ref{ImK_679_nir.fig}\,(e) 
measures the ratio of the flux densities at 2.37\,$\mu$m and 
2.22\,$\mu$m, based on the HST/\-NICMOS filters F222M and F237M:
$$ \rm{CO} = 
 -2.5 \log [\rm{F}_{\lambda} \rm{(F237M)} / \rm{F}_{\lambda} \rm{(F222M)} ]
 - \rm{CO} (\rm{Vega})$$
The correlation is poor, because stronger CO band strengths occur
in redder stars and these continuum changes
nearly compensate for the increasing molecular absorption. 
The index shown in Fig.\,\ref{ImK_679_nir.fig}\,(f) includes
a colour correction: in the above formula,
$\rm{F}_{\lambda} \rm{(F237M)}$ is replaced by the linear 
extrapolation to 2.37\,$\mu$m of the flux densities measured
in the filters F165M (1.65\,$\mu$m) and F222M (2.22\,$\mu$m).
With this redefinition, the flux in the CO band is 
compared to an estimate of the continuum flux at the same
wavelength, rather than to a measurement in an offset continuum filter.
A strong correlation with (I-K) becomes apparent. In fact, the
redefined CO index combines the information contained in the standard
CO index with the information contained in the colour (H-K),
providing a better correlation with (I-K) than any of these two
indices separately (the new linear correlation coefficient is 0.82; 
it was 0.57 for the standard CO index and 0.77 for H-K).

For comparison, we note that the LW2000 sample contains static giants with
(I-K) in the range 2--3.1 and supergiants in the range 2.7-4.5.
For all these objects, the H$_2$O index of Fig.\,\ref{ImK_679_nir.fig}\,(d)
remains below 0.29. The CO indices of giants lie along the mean LPV lines in
Figs.\,\ref{ImK_679_nir.fig}\,(e) and (f), while those of supergiants
lie dispersed along the upper envelope of the LPV distributions. 
\medskip

In view of the large spread of near-IR properties found among O-rich
LPVs, it came as a surprise that the sequence of average 
spectra described in Sect.\,\ref{O_sequence.sec} displayed such
a regular behaviour. Regularity is essential for population
synthesis applications. The effects of pulsation on the upper
atmospheres of LPVs are responsible for a large scatter in the properties
of individual spectra, but do not completely wash out the underlying global 
effect of temperature. The sample of LW2000, for the first time,
provides enough data to recover these trends.

\subsubsection{Secondary parameters for O-rich LPV spectra}
\label{PdV_correl.sec}

As temperature decreases along the TP-AGB, luminosities
rise, radii increase, and masses decrease due to
mass loss. As a consequence, stellar pulsation properties change.
Because of the effects of initial mass, of thermal
pulses and of LPV pulsation, stars with different pulsation
properties can be found at a given TP-AGB temperature.
Essential near-IR features such as the H$_2$O bands depend
on pulsation, even though they also correlate with effective
temperature. In evolving stellar populations, the average
luminosity of TP-AGB stars decreases with age and
a decrease of the average period is thus expected from
empirical period-luminosity correlations (e.g. Feast et al. \cite{FGWC89},
Wood \cite{Wo90}, Wood et al. \cite{WMACHO99}). This
example illustrates that direct effects of pulsation 
properties on spectra at a given temperature will have consequences on
population synthesis predictions.
In principle, population synthesis models should include a gradual
evolution of the pulsation properties instead of simply
considering a dichotomy between variable and non-variable
TP-AGB stars.

We have investigated whether empirical 
correlations in the LW2000 data between spectrophotmetric
properties and pulsation characteristics justify the use
of pulsation criteria rather than temperature in the definition
of averaging bins. Correlations would have to
be significantly better than those of Fig.\,\ref{ImK_679_nir.fig},
to compensate for the fact the theoretical predictions
for the evolution of period and amplitude along the TP-AGB
(or their empirical determination) are even more uncertain than 
temperature predictions and are not included in current
population synthesis codes. As the conclusion is
negative, we only summarize the main empirical results here.

\begin{figure}
\includegraphics[clip=,width=0.5\textwidth]{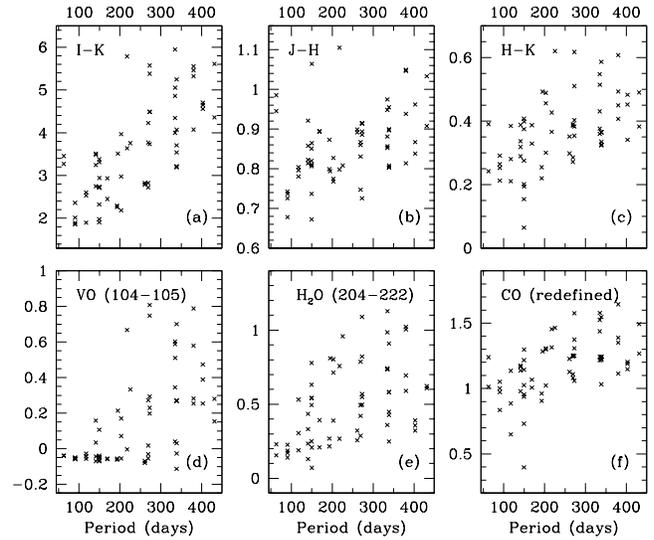}
\caption[]{Colours and near-IR molecular indices versus pulsation
period, for O-rich LPVs.}
\label{P_correlation.fig}
\end{figure}

In Fig.\,\ref{P_correlation.fig}, selected indices
relevant to stellar population studies are plotted against period.
Correlations are clearly present in all panels.
The molecular indices show a larger dispersion at a given period
than they do within the colour-temperature bins of Fig.\,\ref{ImK_679_nir.fig}.
The spread in (J-H) and (H-K) versus period is comparable to the spread seen
in Fig.\,\ref{ImK_679_nir.fig}. We note that no systematic
correlations are found between the residuals in Fig.\,\ref{ImK_679_nir.fig}
and period: for a given spectrophotometric index, the value
and sign of the correlation coefficient changes randomly
from one temperature bin to the next. This probably results
from poor sampling of intrinsically dispersed relations.

\begin{figure}
\includegraphics[clip=,width=0.5\textwidth]{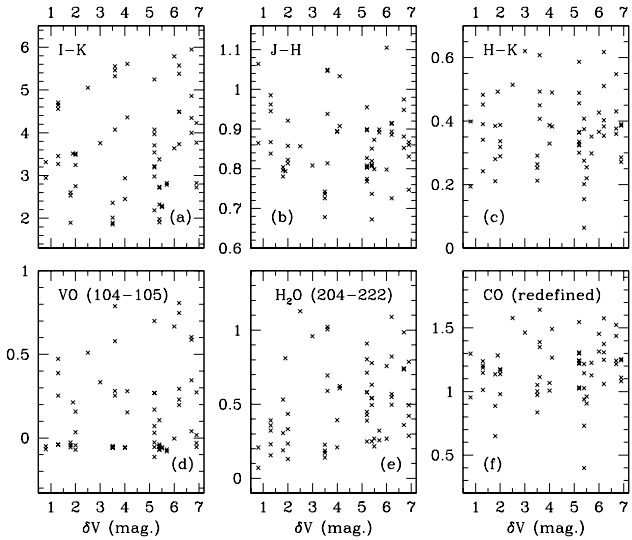}
\caption[]{Colours and near-IR molecular indices versus optical pulsation 
amplitude, for O-rich LPVs.}
\label{dV_correlation.fig}
\end{figure}

Figure\,\ref{dV_correlation.fig} is the equivalent of 
Fig.\,\ref{P_correlation.fig} for amplitude. 
Correlations between amplitude and spectrophotometric indices 
in the sample have a low level of significance, except
for H$_2$O (the correlation coefficient is 0.3, which
for 63 data points is significant at the 98\,\% level;
other correlation coefficients are below 0.15).
Both strong and weak molecular features are found at all amplitudes.
We recall however that the smallest amplitude in the sample is of
0.8\,magnitudes and that no strong bands of water vapour or VO
are seen in the spectra of the static or quasi-static giant
star spectra of LW2000. Again, no systematic correlations were found between
the residuals of the relations of Fig.\,\ref{ImK_679_nir.fig}
and pulsation amplitude.

With the available sample,
the selection of temperature as a primary
sorting criterion for the spectra remains the most sensible
choice. More stellar data are necessary to correctly
define the relations between pulsation properties and 
spectra within individual temperature bins.  

\subsection{Carbon-rich LPVs}
\label{C_sequence.sec}

Depending on initial mass and metallicity, some TP-AGB stars will
become carbon stars through convective dredge-up
of freshly synthesized carbon from the core, induced by thermal
pulses (e.g. Iben \& Renzini \cite{IR83}). Carbon stars are essential
contributors to the near-IR emission from AGB stars, in particular at
subsolar metallicities (Persson et al. \cite{PACFM83}, 
Frogel et al. \cite{FMB90}).
The spectral library contains 21 complete spectra of carbon-rich LPVs.
On the basis of theoretical spectra, Loidl et al. (\cite{LLJ01}, hereafter
LLJ2001) showed that
(R-J) and (R-H) are among the best effective temperature indicators
for these stars. We will (arbitrarily) use (R-H) here.
The far-IR source R\,Lep is so red that it must be treated separately.
R\,Lep excluded, the range of (R-H) values in the sample 
does not justify more than two or three averaging bins.

\begin{figure}[hbt]
\includegraphics[clip=,width=0.5\textwidth]{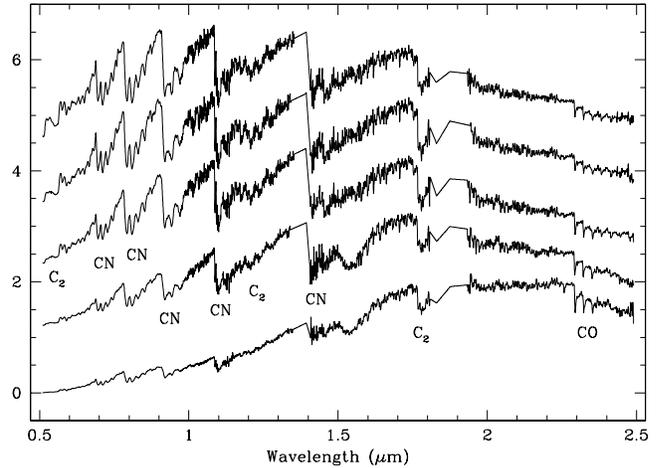}
\caption[]{A temperature sequence of carbon star spectra.
Linear flux density units are used (F$_{\lambda}$, after normalization
to a common total flux, with offsets of 1.1). 
In the regions irremediably affected by telluric 
absorption, around 1.35 and 1.9\,$\mu$m, we have drawn segments by hand
that approximately follow the theoretical shape of molecular bands
(LLJ2001). The carrier of the
molecular band around 1.55\,$\mu$m is uncertain 
(Gautschy \cite{Loi01} suggests a combination of C$_2$H$_2$
and HCN).}
\label{Cseq.fig}
\end{figure}

\begin{figure}[hbt]
\includegraphics[clip=,width=0.5\textwidth]{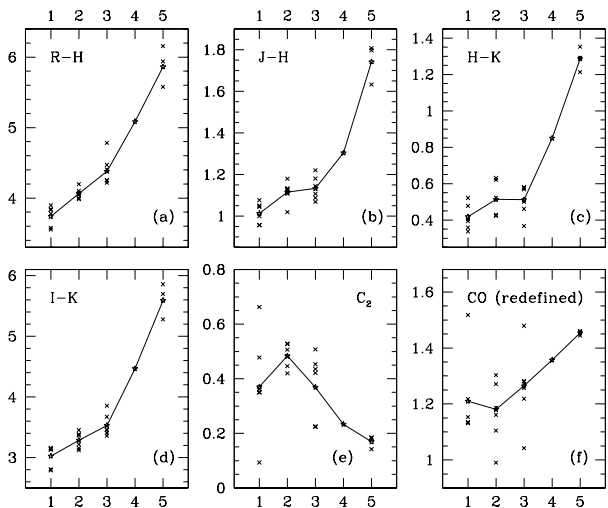}
\caption[]{Spectrophotometric properties of carbon stars
as a function of the bin number based on (R-H). The indices
for instantaneous spectra are shown as crosses; the solid
lines connect those of the averaged spectra (bin 4 contains only 1 spectrum).}
\label{Cstars.photom.fig}
\end{figure}

The suggested temperature sequence for C-rich stars 
contains 3 averages of 6 spectra each (see Table\,\ref{RmH6bins.tab}), 
followed by the individual spectrum of R\,Lep near maximum light, and finally 
the average of the 3 R\,Lep spectra near minimum light (the optical spectrum
in the 4th bin is an average of bins 3 and 5, because no data was obtained).
It is presented in Fig.\,\ref{Cseq.fig}. The averaging procedure
is as for O-rich LPVs. The spectrophotometric
properties of the data bins are shown in Fig.\,\ref{Cstars.photom.fig}.
The C$_2$ index measures the strength of the bandhead at 1.77\,$\mu$m.
To avoid the spectral region of strong telluric absorption corrections
and of contamination by CN, the ``continuum" flux
density is measured over the narrow range 1.752--1.762\,$\mu$m and
the band flux density over 1.768--1.782\,$\mu$m (Alvarez et al. 
\cite{ALPW00}); the index is given in magnitudes. 
The models of Gautschy (\cite{Loi01}) demonstrate
that the C$_2$ bandhead is sensitive to the C/O ratio, 
the $^{12}$C/$^{13}$C ratio and microturbulence. This 
contributes to the dispersion in panel (e). 
More data is needed to assess the significance of the trend 
in panels (e) and (f).

\begin{table}[hbt]
\caption[]{Contents of the C-rich temperature bins}
\label{RmH6bins.tab}
\begin{center}
\begin{tabular}{crllccc} \hline
Bin & Name & Date & C/O$^{a}$ & Period & $\delta$V & (R-H) \\
(1) & (2) & (3) & (4) & (5) & (6) & (7) \\ \hline
1 & T\,Cae & Dec.\,95 & 1.05 & 156. & 1.8 & 3.55 \\
1 & T\,Cae & Jan.\,96 & 1.05 & 156. & 1.8 & 3.58 \\
1 & S\,Cen & Jan.\,96 & 1.40 & 65. & 1.5 & 3.78 \\
1 & T\,Cae & Mar.\,96 & 1.05 & 156. & 1.8 & 3.83 \\
1 & BH\,Cru & Jan.\,96 & 1.01 & 421 & 2.8 & 3.84 \\
1 & Y\,Hya & Dec.\,95 & 1.40 & 302.8 & 3.7 & 3.89 \\
2 & RU\,Pup & Dec.\,95 & 1.10 & 425. & 1.9 & 3.98 \\
2 & RU\,Pup & Jun.\,95 & 1.10 & 425. & 1.9 & 4.00 \\
2 & Y\,Hya & May\,96 & 1.40 & 302.8 & 3.7 & 4.04 \\
2 & Y\,Hya & Jan.\,96 & 1.40 & 302.8 & 3.7 & 4.07 \\
2 & Y\,Hya & Mar.\,96 & 1.40 & 302.8 & 3.7 & 4.10 \\
2 & Y\,Hya & Jul.\,96 & 1.40 & 302.8 & 3.7 & 4.20 \\
3 & RU\,Pup & Jan.\,96 & 1.10 & 425. & 1.9 & 4.22 \\
3 & RU\,Pup & Mar.\,96 & 1.10 & 425. & 1.9 & 4.25 \\
3 & Y\,Hya & Jun.\,95 & 1.40 & 302.8 & 3.7 & 4.26 \\
3 & BH\,Cru & May\,96 & 1.01 & 421. & 2.8 & 4.40 \\
3 & RU\,Pup & May\,96 & 1.10 & 425. & 1.9 & 4.47 \\
3 & BH\,Cru & May\,96 & 1.01 & 421. & 2.8 & 4.78 \\
4 & R\,Lep & Nov.\,97 & --- & 427.1 & 6.2 & 5.08 \\
5 & R\,Lep & Dec.\,95 & --- & 427.1 & 6.2 & 5.58 \\ 
5 & R\,Lep & Jan.\,96 & --- & 427.1 & 6.2 & 5.94 \\
5 & R\,Lep & Mar.\,96 & --- & 427.1 & 6.2 & 6.16 \\
\hline
\end{tabular}
\end{center}

{\em Note to Table}\,\ref{RmH6bins.tab}\,: \\
$a$: Atmospheric number ratio of carbon to oxygen, when available,
according to the estimates of LLJ2001.
\end{table}

Alternative criteria in the construction of averaging bins for carbon stars 
are C/O or $^{12}$C/$^{13}$C, because their effects on the spectral
signatures are significant. Both vary along the TP-AGB
as a consequence of third dredge-up episodes and
envelope burning, in a way that is becoming predictable 
(Renzini \& Voli \cite{RV81},
Forestini \& Charbonnel \cite{FC97}, Marigo \cite{Mar01}). 
Binning with respect to C/O can be done on the basis of estimates of
LLJ2001. It is best to isolate the S/C star BH\,Cru in one bin, 
to combine T\,Cae and RU\,Pup data, and to group Y\,Hya and S\,Cen. 
Unfortunately, current atmosphere models
do not allow us to estimate the ratio for stars as cool as R\,Lep. 
The empirical sequence of increasing C/O ratios is characterised by 
increasing C$_2$ band depths
(at 1.77\,$\mu$m, around 1.2$\mu$m and  below 5700\,\AA ), and by decreasing 
apparent strengths of the CO bandheads around 1.6\,$\mu$m and
2.3\,$\mu$m as a result of heavier contamination by CN and C$_2$ 
(cf. Fig.\,4 of LLJ2001). The C/O and $^{12}$C/$^{13}$C
sequences based on the LLJ2001 estimates are similar (though 
in reverse order).
We note however that a degeneracy with the effects of turbulent velocity
makes $^{12}$C/$^{13}$C estimates uncertain.
The inverse correlation between C/O and $^{12}$C/$^{13}$C found
in the LW2000 sample, and values of $^{12}$C/$^{13}$C as low as 3.5 are 
not typical from the stellar evolution point of view.

Pulsation has small effects on the spectrum below 2.5\,$\mu$m, 
except for the coolest carbon stars, where it allows a deep
band to form around 1.55\,$\mu$m (Fig.\,\ref{Cseq.fig} and
Gautschy \cite{Loi01}).


\subsection{Dust-enshrouded LPVs}
\label{OHIRstars.sec}

OH/IR stars and their carbon rich counterparts are produced
on the upper AGB when the dust produced in the heavy winds
of an LPV builds a thick enough envelope to hide the 
central star at optical wavelengths. At a given bolometric
luminosity, their contribution to the near-IR light 
is smaller than for unobscured stars but their mid-IR
emission is significant, an effect that may be useful in
searches for intermediate age stellar populations
(Bressan et al. \cite{BGS98}). More than the effective temperature
of the central star, the amount of dust (which is related
to the mass loss and, indirectly, also to the effective
temperature) determines the near-IR energy output.

\begin{figure}
\includegraphics[clip=,width=0.5\textwidth]{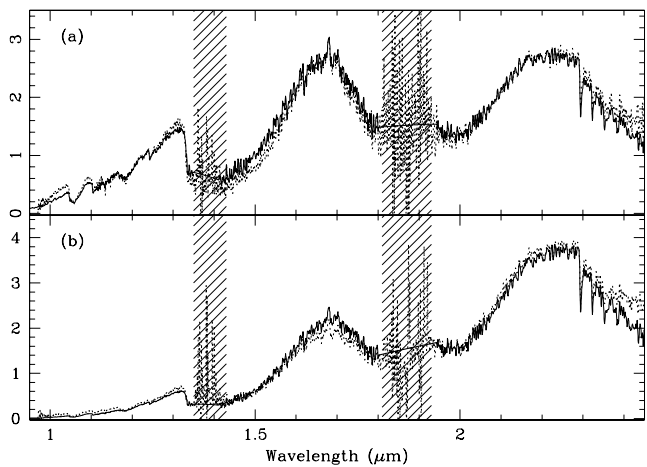}
\caption[]{{\bf (a)}: Average of the April 1995 and May 1996 spectra
of AFGL\,1686\,=\,IRAS 14068-0730 (dotted) and reddened version
of the spectrum of bin\,9 of Sect.\,\ref{O_sequence.sec}, using the
Milky Way extinction law with Av=3 (solid). {\bf (b)}:
Spectrum of WX\,Psc\,=\,IRAS 01037+1219 (dotted) and reddened
spectrum of bin\,9 with Av=18 (solid). F$_{\lambda}$ units
are arbitrary.}
\label{ext_law.fig}
\end{figure}

To represent the optical and near-IR emission of dust-enshrouded 
sources, we suggest to use reddened versions of the average spectra 
provided in this paper. Although circumstellar dust might
be of a different nature than the average interstellar
dust grains, the LW2000 data suggest that the Milky Way
extinction law (e.g. Cardelli et al. \cite{CCM89}, with 
Av/E(B-V)=3.1) can be used to first order for that purpose.
The library of LW2000 contains a handful of spectra of OH/IR
sources. In Fig.\,\ref{ext_law.fig}, AFGL1686 and
WX\,Psc are compared with reddened versions of the coolest
average spectrum of Sect.\,\ref{O_sequence.sec}, which
has similar bands of H$_2$O, TiO and VO. The agreement
is rather good (and even better agreements can be found with 
reddened versions of selected individual spectra of the initial
spectral library). The Milky Way extinction law creates
no obvious discrepancy. More data will be needed to 
investigate this question further, and to perform
similar tests with carbon rich stars.

\section{Using the spectra in population synthesis codes}
\label{useit.sec}

For practical use in connection with theoretical stellar evolution
tracks, the empirical spectra must be properly normalized and
a quantitative effective temperature scale must be adopted.
These aspects are discussed in Sect.\,\ref{bolcor.sec} and
Sect.\,\ref{Tscales.sec} below.
Of course, the stellar evolution tracks must also provide 
the transitions between O-rich and C-rich atmos\-pheres, and 
between optically visible sources and dust-enshrouded infrared sources.
The discussion of these aspects of theoretical stellar evolution lies
outside the scope of this paper.

\subsection{Bolometric corrections}
\label{bolcor.sec}
\begin{figure}[!ht]
\includegraphics[clip=,width=0.5\textwidth]{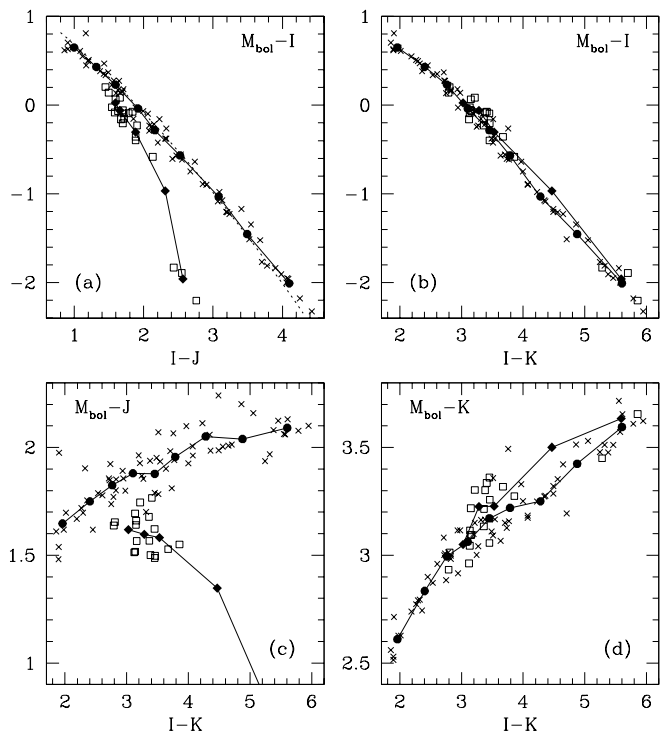}
\caption[]{Bolometric corrections, obtained from the LW2000 data and
the mean spectra, with the extrapolation method described in the text.
Crosses and open squares: individual stellar spectra for O-rich and
C-rich LPVs; solid symbols: sequences of averaged spectra. Dotted line
in {\bf (a)}: values obtained by Alvarez et al. (\cite{ALPW00}).
}
\label{Bcors.fig}
\end{figure}

Bolometric corrections describe the relation
between the spectral flux densities and the bolometric luminosity
L$_{\rm bol}$.
As the LW2000 data encompass most of energy emitted by the LPVs,
only small corrections are necessary to convert the  
measured integrated flux into a bolometric flux. For
oxygen rich stars, Alvarez et al. (\cite{ALPW00}) 
computed these corrections by extrapolating
the LW2000 data with the best-fitting cool M giant models. 
They found that on average about 20\,\% of the light is emitted 
outside the range of the available data. In practice,
because blackbodies are poor representations of the infrared
emission of cool stars (Fluks et al. \cite{FPTetal94},
Aringer et al. \cite{AKHetal99},
J{\o}rgensen et al. \cite{JJSA01}), 
we suggest to simply extrapolate the data linearly to a nil
flux at 3500\,\AA\ on the blue side and to 50\,000\,\AA\ on the
red side. These extrapolation limits are chosen to reproduce
the results of Alvarez et al., as shown in Fig.\,\ref{Bcors.fig}\,(a).
The long wavelength extrapolation also provides agreement
with the integrated fluxes measured longward of 2.5$\mu$m 
for the six ISO satellite spectra of oxygen 
rich LPVs we have examined\footnote{SV\,Peg, T\,Sge, V1351\,Cyg, V584\,Apl, 
VY\,Cas and CE\,And; available from the public domain page of Kerschbaum,
Loidl, Hron, Kerber, and Rauch: 
{\em www.astro.univie.ac.at/}\,$\sim${\em fzi/AGB/agb2pn.html}.},
to within random deviations smaller than 10\,\% (i.e. a few percent on
L$_{\rm bol}$).

The infrared emission of carbon stars also differs
significantly from a blackbody (e.g. Goebel
et al. \cite{GBSWE78}, Aoki et al. \cite{ATO99}). 
According to a subset of the models described in LLJ2001, and found to provide 
satisfactory agreement with ISO data (J{\o}rgensen et al. \cite{JHL00}), 
the wavelength range of the LW2000 data
again comprises 80\,\% of the flux, and only a few percent
of the missed flux comes from $\lambda<5000\,\AA$. 
The total flux of ISO spectra\footnote{TX\,Psc, V460\,Cyg and
S\,Cep; retrieved from the Vienna web
site mentioned above.} longward of 2.5\,$\mu$m is
equivalent to the suggested linear extrapolation, with 
object-to-object differences that amount to variations
of up to 10\,\% in the total fluxes. The absence of 
ISO counterparts for the LW2000 stars makes it impossible
to test directly whether the extrapolation is still valid
for the coolest carbon star bin, but the near-IR similarity 
between R\,Lep and the ISO target S\,Cep (see Goebel et al.
\cite{GBSWE78} and LW2000) suggests that the linear extrapolation
method doesn't underestimate the L$_{\rm bol}$ by more
than 30\,\% even in that case. 


Figure \ref{Bcors.fig} shows that the bolometric corrections 
M$_{\rm bol}-$I and M$_{\rm bol}-$J for carbon stars lie below  
those of oxygen rich stars. The same would be true for M$_{\rm bol}-$H.
The explanation lies in the location of the main molecular bands.
In O-rich stars, water bands tend to lie at the edge of the
standard photometric windows, while in carbon stars the C$_2$ and
CN bands reduce the flux density right through these windows.
In comparison, molecular absorption is relatively small in the K window
for both chemical classes.

The bolometric corrections above do not apply to AGB stars
in opaque dust shells. Applying reddening as suggested in  
Sect.\,\ref{OHIRstars.sec} automatically accounts for the reduced
optical and near-IR fluxes of these objects. The determination
of bolometric corrections is thus directly dependent on 
the rapidly evolving circumstellar optical depth.

\subsection{Temperature scales}
\label{Tscales.sec}

The second step required in the assignment of a spectrum to
a particular stage along a stellar evolution track is the 
determination of an effective temperature (T$_{\rm eff}$) scale 
for the available spectra. A straightforward way of doing this is to rely 
on theoretical or semi-empirical scales (e.g. angular diameter measurements).
Although we provide relations of that kind here, we 
emphasize that with current stellar models all attempts
to relate temperatures assigned this way with 
the T$_{\rm eff}$ of a particular set of stellar evolution tracks
are uncertain. More arguments to this important point 
are given in Sect.\,\ref{hr_motions.sec}.

Bessell et al. (\cite{BBSW89b}) provide
temperature scales based on (I-K).
The scales are metallicity-dependent. 
We adopt solar metallicity here and refer to Sect.\,\ref{metal.sec} 
for a discussion of metallicity effects. 
Synthetic spectra for static and pulsating AGB stars have been kindly
made available to us by M.\,Scholz. The static models and their colours
are described in Bessell et al. (\cite{BBSW89b}, \cite{BBSW91}), 
the pulsating ones in Bessell et al. (\cite{BSW96}; 
their E, Z and D series) and Hofmann et al. (\cite{HSW98};
their L, M, P and Q series). A detailed comparison
between the models and the data lies far beyond the scope of this paper.
Although many observed tendencies are reproduced by
the models at least in a statistical sense, the fits to individual
spectra are not perfect, and from our preliminary analysis
we estimate that they do not provide a more 
reliable absolute effective temperature than, for instance, a colour
such as (I-K). 

\begin{figure}[!hbt]
\includegraphics[clip=,width=0.5\textwidth]{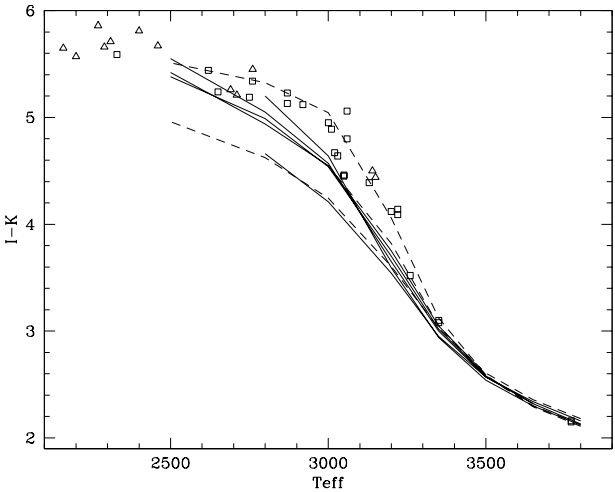}
\caption[]{Effective temperature scales based on synthetic spectra.
The lines follow static model series. Dashed  (Bessell et al. \cite{BBSW89b}): 
1\,M$_{\odot}$ sequences with, from top to bottom 
(10$^4$\,L$_{\odot}$; Z$_{\odot}$), (2\,10$^3$\,L$_{\odot}$; Z$_{\odot}$),
(10$^4$\,L$_{\odot}$; Z$_{\odot}/2$).
Solid (Bessell et al. \cite{BBSW91}): the bundle of 4 similar lines follow 
the AGB locus at Z$_{\odot}$ for 1, 1.5, 2.5 and 5\,M$_{\odot}$,
and the bluer line is for 1\,M$_{\odot}$ at Z$_{\odot}/2$. 
Symbols (Bessell et al. \cite{BSW96}, Hofmann et al. \cite{HSW98}):  
dynamical model stars at phase 0.5 (triangles) or 0.8 or 0. (squares).
}
\label{Teff_ImK_stat_dyn.fig}
\end{figure}

Figure\,\ref{Teff_ImK_stat_dyn.fig} shows the relation between
(I-K) and the effective temperature 
for solar metallicity models and for two series of static models
at subsolar metallicity. The effective temperature is a theoretical quantity,
defined using Stefan's law and 
the radius at which the Rosseland mean optical depth of the star reaches unity. 
While each of the dashed lines is a sequence of models 
at constant mass (1\,M$_{\odot}$) and luminosity, the solid lines 
follow the luminosity-temperature relations of AGB evolutionary tracks 
as parametrized by Bessell et al. (\cite{BBSW91}). The symbols shown for the 
pulsating models are instantaneous values, corresponding to
a particular cycle and pulsation phase. It is seen that the dynamical
models follow a sequence that agrees with the 1\,M$_{\odot}$,
high luminosity (i.e. low gravity) static solar metallicity 
sequence of Bessell et al. (\cite{BBSW89b}). This locus
provides the temperatures of the averaged empirical spectra listed
in Col. 5 of Table\,\ref{prop_av.tab}. A lower effective temperature would be
assigned to a given spectrum if the subsolar theoretical scales or
the solar ones of Bessell et al. (\cite{BBSW91}) were used instead.
The comparison of the range in (I-K) in Table\,\ref{prop_av.tab}
and along the models in Fig.\,\ref{Teff_ImK_stat_dyn.fig} shows
that the sequence of average LPV spectra adequately covers the
temperature range required for AGB synthesis.

\begin{table}
\begin{center}
\caption[]{Properties of the averaged spectra}
\label{prop_av.tab}
\begin{tabular}{ccccc} \hline
\mcol{4}{c}{O-rich spectra} \\ \hline
Bin & I-K & J-K (J-Ks)$^{a}$ & BC(I) & T \\
(1) & (2) & (3)  & (4) & (5) \\ \hline 
1 & 1.96 & 0.96 (1.01) & 0.65 &  3930\,K \\
2 & 2.40 & 1.08 (1.13) & 0.43 &  3585\,K \\
3 & 2.76 & 1.17 (1.22) & 0.23 &  3440\,K \\
4 & 3.10 & 1.18 (1.23) & -0.04 &  3355\,K \\
5 & 3.45 & 1.29 (1.34) & -0.28 & 3285\,K \\
6 & 3.79 & 1.26 (1.31) & -0.57 & 3235\,K \\
7 & 4.28 & 1.20 (1.24) & -1.03 & 3175\,K \\ 
8 & 4.88 & 1.38 (1.43) & -1.45 & 3055\,K \\
9 & 5.60 & 1.50 (1.55) & -2.01 & 2340\,K \\ \hline
\mcol{4}{c}{C-rich spectra} \\ \hline
Bin & R-H & J-K (J-Ks)$^a$ & BC(I) & T \\
(1) & (2) & (3)  & (4) & (5) \\ \hline
1 & 3.74 & 1.43 (1.52) & 0.02 & 3200\,K \\
2 & 4.06 & 1.63 (1.71) & -0.06 & 3000\,K\\
3 & 4.38 & 1.65 (1.74) & -0.30 & 2800\,K\\
4 & 5.08 & 2.15 (2.23) & -0.97: & 2400\,K\\
5 & 5.86 & 3.03 (3.08) & -1.96: & 2000\,K\\ \hline
\end{tabular} 
\end{center}

{\em Notes to Table}\,\ref{prop_av.tab}\,: \\
$a$\,: (J-Ks) is given for the DENIS filter passbands (Epchtein et al.
\cite{Epetal97}), kindly provided by M.\,Schultheis.\\
A colon indicates uncertain values, where the far-IR emission
is likely to be underestimated by the adopted extrapolation procedure.\\
For the interpretation of Col.\,5, see Sect.\,\ref{Tscales.sec}
and \ref{hr_motions.sec}.
\end{table}

\medskip

For carbon stars with effective temperatures above $\sim 2600$\,K, 
we adopt the relation between (R-H) and T$_{\rm eff}$
derived by LLJ2001. The values for the 
binned spectra of Fig.\,\ref{Cseq.fig} are given in Table\,\ref{prop_av.tab}. 
The two coolest bins of the sequence contain spectra of R\,Lep,
a Mira-type variable with a period of 427 days.
A T$_{\rm eff}$ of $2058 \pm 180$\,K has been obtained for this star
from interferometric angular size measurements in October, 1995 (van
Belle et al., \cite{vBetal97}). The data in our temperature bin\,5 were taken
between December, 1995 and early March, 1996.  The 3 individual 
spectra are very similar; the comparison of their energy distributions
suggests that the minimum temperature was reached in early 
January, 1996, i.e. only 0.2 cycles after the interferometric measurement. 
We therefore assign the spectrum of bin\,5 an effective temperature
of $\sim 2000$\,K. The (V-K)--T$_{\rm eff}$ relation for Carbon Miras
of van Belle et al. (\cite{vBetal97}) predicts (V-K)\,$\simeq$\,10 at this 
temperature, which is in good agreement with our measurement (after
extrapolation of the observed spectrum to the short wavelength
cut-off of the standard V band filters). Based on this sequence,
T$_{\rm eff} \simeq 2400$\,K is appropriate for bin\,4.
The range of (J-K) values in the average spectra is in excellent agreement
with observations in complete carbon star samples 
(Frogel \& Elias \cite{FE88}, Glass et al. \cite{GWCF95},
Cioni et al. \cite{CvMLH00}, Weinberg \& Nikolaev \cite{WN01}).

\section{Discussion}
\label{disc.sec}

\subsection{Temperature bins and temperature calibrations}
\label{hr_motions.sec}

In the above sections, we have shown that wide baseline colour
temperatures provide the most sensible and practical classification
of the LPV spectra. When using these binned spectra to synthesize cluster
or galaxy spectra, 
one assumes that the energy-weighted mean spectrum of an
individual LPV (
averaged over the $\leq 10^3$\,days of the LPV pulsation cycle)
is similar to the average of many individual
stellar spectra obtained from stars of various masses,
amplitudes and phases, but with a common colour temperature.
As we found no significant correlations, in any of our temperature
bins, between the deviations from the mean spectral properties and
parameters such as amplitude or period, this is a reasonable choice.
More observations will be needed before other options can be considered. 

With  the above assumption, the
effective temperature along a TP-AGB evolution track alone determines 
which spectrum should be used to represent the corresponding emission
(it is assumed that the track also indicates when the use of LPV spectra instead
of static giants becomes necessary, and when a star becomes C-rich
or dust-enshrouded). In order to avoid misinterpretations of the
temperature scales given in Table\,\ref{prop_av.tab}, it is important 
to recall how evolutionary TP-AGB tracks and individual TP-AGB stars
behave in the HR diagram.
\medskip

\begin{figure}[!ht]
\includegraphics[clip=,width=0.5\textwidth]{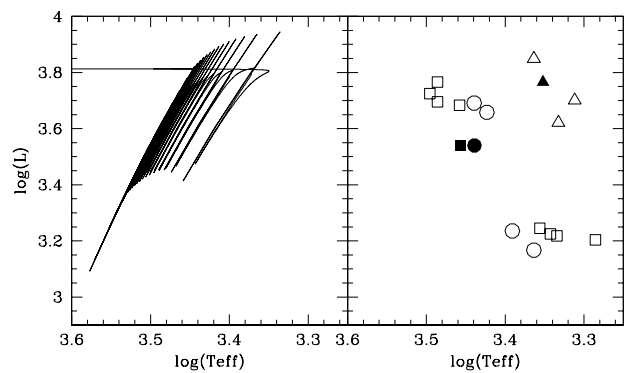}
\caption[]{{\bf Left}\,: Synthetic TP-AGB track for the evolution of a star 
with an initial mass of 2\,M$_{\odot}$ and a final mass of 
0.6\,M$_{\odot}$.
{\bf Right}\,: 1\,M$_{\odot}$ pulsating star models from Hofmann et al. 
(\cite{HSW98}).
The P, M and O series are shown respectively as squares, circles and 
triangles. Properties at maximum and minimum light in several successive
cycles are shown (open symbols), together with those of the static parent star
(solid).
}
\label{theoryHR.fig}
\end{figure}

\begin{figure}[htb]
\includegraphics[clip=,width=0.5\textwidth]{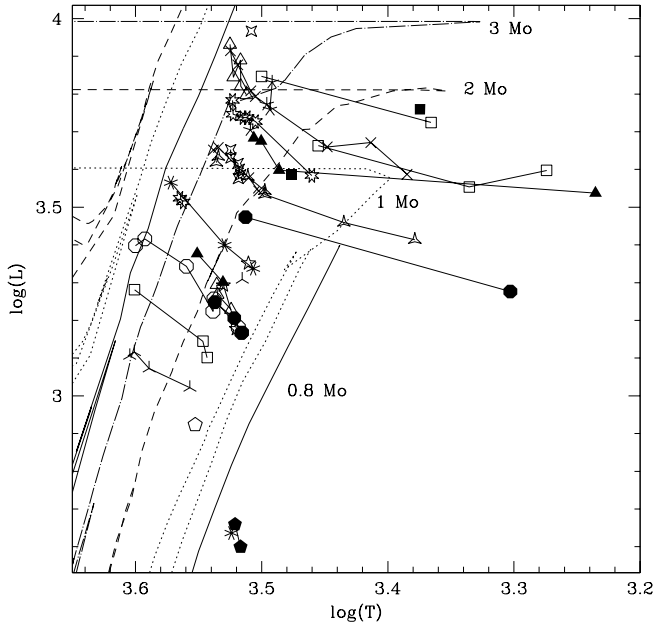}
\caption[]{Tentative location of the LW2000 spectra on the theoretical
HR diagram (for stars with known periods). 
\Teff\ is derived from (I-K), log(L) from the period and the instantaneous
bolometric correction. Symbols distinguish individual stars; multiple
observations at various phases are connected. Smoothed AGB tracks
at Z$_{\odot}$ are overlaid (Mouhcine \& Lan\c{c}on \cite{ML02_models};
at 0.8\,M$_{\odot}$, only the RGB is shown).
}
\label{Stars_on_HR_MM.fig}
\end{figure}

Even a perfectly coeval stellar population produces a broad asymptotic
giant branch in the HR diagram and in colour-magnitude diagrams,
as a result of the thermal pulses (helium shell flashes) that occur
on timescales of $\sim 10^4$\,yrs and of the long period pulsational
variability on timescales of $10^2-10^3$\,days. Figure 
\ref{theoryHR.fig} shows theoretical
illustrations of these effects, and Fig.\,\ref{Stars_on_HR_MM.fig} 
provides an empirical counterpart.

The TP-AGB track on Fig.\,\ref{theoryHR.fig} (left) 
was computed using so-called ``synthetic evolution" techniques 
(Iben \& Renzini \cite{IR83},  Groenewegen \& de Jong \cite{GdJ93},
Wagenhuber \& Groenewegen \cite{WG98}, Marigo et al. \cite{MBC98}, 
Mouhcine \& Lan\c{c}on \cite{ML02_models}). 
The values of the various model parameters
for this particular track are not relevant here. 
Tiny offsets in many stellar parameters will ensure that
thermal pulses do not occur in phase for all stars. Most stars 
will be found around the quiescent luminosity, and $10-30\%$ 
at the significantly lower luminosities that follow thermal 
pulses (Boothroyd \& Sackmann \cite{BS88a}, Groenewegen \& de Jong 
\cite{GdJ93}, Marigo et al. \cite{MGB99}).  The locus of
any set of TP-AGB tracks is based on its own particular definition
of the T$_{\rm eff}$ (see Scholz \& Takeda \cite{ST87}
for a discussion), through rather simple outer boundary conditions
for such cool and extended stellar atmospheres. In addition,
as already mentioned in Sect.\,\ref{intro.sec}, 
these temperatures are those of {\em static}\ model stars.

The right hand plot of
Fig.\,\ref{theoryHR.fig} shows the predicted effect of Mira-type
pulsation for the three most recent models described by Hofmann et al.
(\cite{HSW98}). All models pulsate with periods of the order of 330 days 
and assume a mass of 1\,M$_{\odot}$.
The two models at relatively lower luminosity (series P and M
of the authors) are fundamental mode pulsators, while the third one (O series) 
pulsates in the first overtone. All models behave differently
in the HR diagram. Model M shows that pulsation is able 
to shift a star {\em systematically} to larger radii and
lower effective temperatures, when compared to its static parent. 
This trend is also seen when direct angular diameter measurements of
pulsating and static late-type giants are compared (van Belle et al.
\cite{vBetal97}). However, the amplitude of this effect at
a given point of an evolutionary track cannot be predicted
reliably because neither the evolution of LPV pulsation modes,
nor pulsation amplitudes, nor the actual effect of these
quantities on the spectra are understood.
\medskip

In Fig.\,\ref{Stars_on_HR_MM.fig}, the O-rich
stars of the LW2000 library have been located tentatively
in the theoretical HR diagram, in order to estimate the empirical
width of an upper AGB sequence. Each star with a known period has been
assigned a K band luminosity (L$_{\rm K}$) using the 
empirically well-defined period -- L$_{\rm K}$ relation for Miras
by Hughes \& Wood (\cite{HW90}). Note that those stars which pulsate in 
overtone modes should in fact be assigned higher luminosities (see
LW2000 for potential candidates).
L$_{\rm K}$ variations have been neglected in the figure.
The changes in the bolometric luminosity represent
the changes in the instantaneous bolometric corrections 
measured on the spectra as in Sect.\,\ref{bolcor.sec}. The assigned 
temperatures are derived from (I-K) as in Sect.\,\ref{Tscales.sec}. 
It is seen that pulsation can widen the
AGB to $\Delta {\rm T} = 500-1000\,\rm{K}$ or
$\Delta {\rm (I-K)} = 1-2\,\rm{mag}$. This range is 
consistent with the predicted ranges shown in Fig.\,\ref{theoryHR.fig}.
If pulsation was neglected, such a range of colours could be mistaken
for a range of several M$_{\odot}$ in initial stellar masses. Alternatively,
it could be attributed to a spread in metallicity. TP-AGB tracks are
typically shifted to hotter T$_{\rm eff}$ by 0.3\,dex when 
going from Z=0.02 to Z=0.008.

The tracks overlaid on Fig.\,\ref{Stars_on_HR_MM.fig} are
smoothed (many population synthesis codes require that 
thermal pulses be smoothed out,
to make interpolations between tracks or the construction of
isochrones possible). The smoothed luminosities are the
result of energy conservation: smooth tracks
run at luminosities slightly lower than the quiescent, pre-flash 
luminosities, in order to account for the post-flash luminosity dips. 
More options exist in the smoothing schemes for T$_{\rm eff}$,
as the ideal weighting should use the energy output in the particular 
wavelength range one intends to make predictions for.
The differences between various schemes will however remain
small compared to the systematic effects of T$_{\rm eff}$ definitions or
LPV pulsation mentioned above.
\medskip

In summary, theoretical effective temperatures are 
not directly comparable to those estimated for individual empirical spectra of
LPVs. Acquiring enough data to obtain representative average
spectra is only one step, in which the LW2000 sample has allowed significant
progress. But until consistent models for the structure and spectrum
of pulsating AGB stars exist, the link between the
theoretical T$_{\rm eff}$ scale and any semi-empirical one 
remains uncertain. It must be calibrated a posteriori.
Ideal observational targets for the temperature
calibrations are {\em massive} stellar 
clusters with known metallicities and ages (from optical spectroscopy),
such as found in merger galaxies for instance (Maraston et al. \cite{MKBBH01},
Mouhcine et al. \cite{MLLGS02}). The relevant ages lie between 10$^8$ 
and 2\,10$^9$\,yrs. In individual clusters containing less 
than $\sim 10^5$ stars, the stochastic effects due to the intrinsically 
small number of bright AGB stars affect the comparison significantly 
(Lan\c{c}on \& Mouhcine \cite{LM00_fluct}).

\subsection{Period and amplitude distributions}
As recalled in Sect.\,\ref{sample.sec},
the stars of the LW2000
sample were selected to provide a relatively uniform coverage
of the period--amplitude plane. Those used to  construct 
the averages of this paper cover periods of $100\leq$\,P\,$\leq 450$ 
and optical amplitudes of $0.8\leq \delta {\rm V} \leq 7$\,magnitudes.

Systematic photometric observations of large samples of stars have shown that
the majority of the TP-AGB stars are variable (Wood et al. \cite{WMACHO99},
Alcock et al. \cite{AlcMACHO00}, Cioni et al. \cite{CMLetal01}). 
However, period and amplitude distributions
are expected to depend on the age of the observed population
(Vassiliadis \& Wood \cite{VW93}). Large amplitude pulsation
characterizes only a subpopulation of the TP-AGB stars. 
Here, we discuss whether or not it is appropriate to use
the average LPV spectra of this paper for {\em all} the TP-AGB
stars of a synthetic population.
\medskip

Period distributions in the Large Magellanic Cloud (LMC) or in
the Galaxy extend from less than 20 days to more than $10^3$ days. The
range of periods of our LPV sample corresponds 
to those of optically visible Miras 
and of relatively long period, high luminosity semi-regular stars, as seen
in Fig.\,1 of Wood (\cite{Wo00}). Both in the LMC and in the
Galactic Bulge, semi-regular variables with periods shorter than
100\,days outnumber Miras (Wood et al. \cite{WMACHO99}, 
Alard et al. \cite{AlaMACHO01}, Alcock et al. \cite{AlcMACHO00}).
The mean period and the mean amplitude of our sample are thus larger 
than the mean period of optically
visible TP-AGB stars in populations that have been forming stars for 
a long time (i.e. since the epoch of galaxy formation).
\medskip

What is relevant to the synthesis of {\em integrated} galaxy
or cluster spectra is how the period and amplitude distribution
of the available stellar sample compares with the distribution for those
particular TP-AGB stars that contribute most strongly to the
light (in the near-IR, i.e. the only spectral range of the LW2000 data
where the TP-AGB contribution is significant). 
Population synthesis models have shown that these
are stars with initial masses between about 1.8 and 3\,M$_{\odot}$
(Mouhcine \& Lan\c{c}on \cite{ML02_models}; see also 
Renzini \cite{Ren92}, Girardi \& Bertelli \cite{GB98},
Maraston \cite{Mara98}). They reach the TP-AGB
with ages of about 0.1 to 1.3\,Gyr and spend of the order of $10^6$\,yrs
in that phase (exact ranges are model and metallicity dependent).
In older populations, the stars of the red giant
branch are so numerous that they provide most of the near-IR 
light. 
Which spectra are used for the remaining insignificant TP-AGB
stars stops being critical then.

What do we know about the pulsational properties of stars
with initial masses between 1.8 and 3\,M$_{\odot}$?
A global picture has emerged from joint studies of 
empirical period-luminosity distributions and theoretical
pulsation models (Fox \& Wood \cite{FW82}, Xiong et al. \cite{XDC98},
Vassiliadis \& Wood \cite{VW93}, Wagenhuber \& Tuchman \cite{WT96}).
In the period-luminosity diagram,
a single evolutionary track runs ``stepwise" accross  pulsation mode
sequences, from short to long periods, with a slowly increasing
luminosity (apart from the temporary changes due to the thermal
pulses): a star becomes unstable to one of the overtone pulsation
modes, follows that period-luminosity sequence until it switches to a 
lower overtone, and so on. The large mass loss associated with
Mira-type pulsation finally makes the star loose its envelope.
Stars with more massive initial masses
arrive in the unstable strips of the period-luminosity diagram with
larger luminosities.  This is supported by the
observation that Milky Way Miras with very long periods are 
associated with the thin disk, while shorter period Miras are 
seen in the thick disk and in globular clusters (Feast \cite{Fea63},
Habing \cite{Hab95}). The stars with masses between 1.8 and 3\,M$_{\odot}$
are thus expected to pulsate with larger mean periods than a
population in which most stars are old.

\begin{figure}[!htb]
\includegraphics[clip=,width=0.5\textwidth]{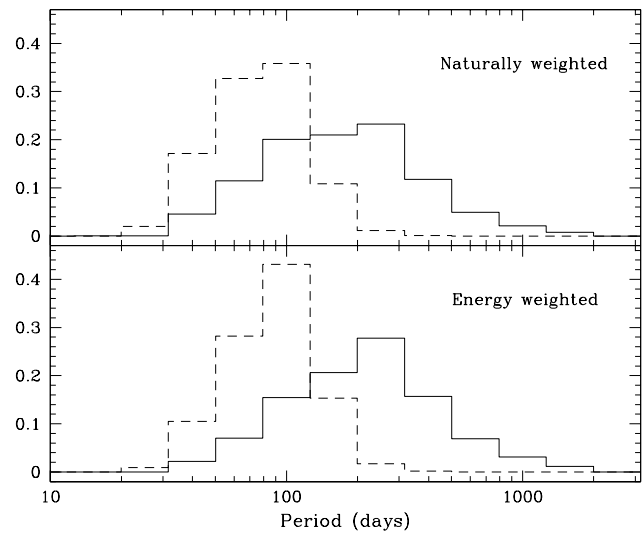}
\caption[]{Predicted period distributions for stars with initial
masses around 2\,M$_{\odot}$ (Z=Z$_{\odot}$). 
Fundamental mode pulsation is assumed
for the solid line, first overtone for the dashed line.
Each distribution is normalized individually to
a sum of one.
}
\label{Pdistrib.fig}
\end{figure}

Figure \ref{Pdistrib.fig} shows predicted period distributions
for a stars with an initial mass of 2\,$M_{\odot}$, 
based on the TP-AGB tracks of Mouhcine \& Lan\c{c}on (\cite{ML02_models}),
combined with the period-mass-radius relations of Wood (\cite{Wo90}) 
for fundamental mode and of Barth\`es \& Luri (\cite{BL01}) for the first
overtone.
Systematic effects of the initial mass are small between 
1.8 and 3\,$M_{\odot}$. The plotted distribution
is representative of the initial masses most relevant to observable integrated 
properties of stellar populations. The upper graph is a number distribution,
the lower gives the relative contributions of stars with
various periods to the luminosity. The relative number of fundamental
mode and overtone pulsators is very uncertain. In the LMC period-luminosity
diagram (Wood et al. \cite{WMACHO99}), about 1/3 of the LPVs in the 
relevant, high luminosity range are assigned fundamental mode
(unfortunately the LMC scarcely populates the upper part of the
diagram, due to the combined effects of the stellar initial mass
function and the star formation history). Fig.\,\ref{Pdistrib.fig}
shows that the range of periods available to us for the construction
of average spectra is indeed very appropriate for the fundamental
mode pulsators. A similar conclusion is reached when other tracks in 
the literature are used (Vassiliadis \& Wood \cite{VW93}, Wagenhuber \& 
Tuchman \cite{WT96}, Gautschy \cite{Gau99}).

The periods of our spectroscopic sample are however larger than
those of the overtone pulsators. 
According to the data of Wood et al. (\cite{WMACHO99}), 
the overtone pulsators are 
essentially all semi-regular variables, with small amplitudes compared to
those of the Mira-like fundamental mode pulsators (semi-regular
LPVs are found on the fundamental mode sequence, but at lower
luminosities and shorter periods than those relevant here,
as seen in the light curves given in Wood et al. \cite{WMACHO99}).
Many of the overtone pulsators should be well represented
with spectra of static giants. Thus, the relative importance of
the LPVs in synthetic populations can be varied by modifying
the blue edge of the instability strip. 

Current pulsation theory 
does not put strong constraints on this edge and on the transition
from overtone to fundamental mode pulsation. In order to evaluate
the effects of such adjustments, we have compared synthetic spectra of 
intermediate age populations computed with two assumptions: (i) our average 
LPV spectra are appropriate for all TP-AGB stars or (ii) the
LPV spectra only apply in the Mira instability strip. For (ii), the blue
edge of the Mira instability strip was projected from 
the empirical period-luminosity plane into the HR diagram
assuming fundamental mode pulsation models (Wood \cite{Wo90}), 
and static giant stars were used up to that point.
The differences were much smaller than the uncertainties
currently associated with, for instance, the effective temperature
scales of the spectra (changes were smaller than 3\,\% in all the
quantities we measured).
\medskip

In summary, the simplifying assumption that the average LPV spectra 
of this paper apply to all TP-AGB stars is reasonable,
when the aim is the synthesis of integrated spectra of stellar populations.
Better theoretical predictions for the onset and evolution of
period and amplitude of TP-AGB stars with various initial 
masses will be needed to reduce the uncertainties inherent
to this assumption. In the study of resolved populations
that include AGB stars with low initial masses, it is 
likely that the spectra of static giant stars should be 
used for some as yet poorly defined first part of the TP-AGB.

\subsection{Hydrogen recombination lines}
The only conspicuous near-IR emission line seen in LPVs is
Paschen $\beta$, at 1.28\,$\mu$m. Emission equivalent widths
of up to 6\,\AA\ are observed in individual spectra of the LW2000 sample.
Pa\,$\beta$ is seen in most of the averaged O-rich spectra of 
Sect.\,\ref{bin_choice.sec}. The emission equivalent width is of 
the order of 1\,\AA\ in 6 of the 9 temperature bins. None of the 
C star spectra of LW2000 shows significant Hydrogen recombination
lines. This is mostly due to molecular line crowding, which is more
severe in C-rich spectra around the wavelengths
of the hydrogen lines. In the maximum light spectrum of the 
S/C type star BH\,Cru, Pa\,$\beta$ is detected with an
equivalent width of the order of 1.5\,\AA\ (the resolution of
the data defines the pseudo-continuum). Paschen $\beta$ is not
seen in the averaged C star spectra of Sect.\,\ref{C_sequence.sec}.

According to the evolutionary tracks and population synthesis
calculations of Mouhcine \& Lan\c{c}on (\cite{ML02_models}), TP-AGB
stars contribute less than 40\,\% of the J band light
in coeval populations of intermediate ages.
If the Pa\,$\beta$ equivalent widths of the average LPV spectra
are representative of the TP-AGB, 
those of integrated spectra of stellar 
populations will not exceed fractions of an Angstr\"om. This is 
small compared to the effect of any significant subpopulation of 
young ionizing stars (Leitherer et al. \cite{LSG99}).

\section{A recipe for spectral synthesis at various metallicities}
\label{metal.sec}

When metallicity is varied, several effects add up to produce
differences in the predicted contributions of upper AGB stars
to the integrated light of a stellar population. 
First, the relation between effective
temperature and spectrophotometric properties changes. For cool
stars, this is due to the importance of molecular opacities. An illustration
can be found, e.g., in Fig.\,10 of Hauschildt et al. (\cite{HAFBA99}),
where static giant model spectra for $[$M/H$]$=0 and -0.3
are compared. When working at lower metallicity, a bluer
spectrum with weaker molecular bands must be associated with each given point 
of the theoretical HR diagram. Second, the AGB evolutionary tracks 
themselves are globally shifted to lower effective temperatures. 
This effect acts in the same direction as the first one: signatures
of cooler AGB stars will be more pronounced at higher metallicities.
The third effect is the metallicity-dependence of the mass loss,
of the AGB lifetimes and of the efficiency of the production of carbon stars.
Discussions of these latter aspects can be found, e.g., in
Groenewegen \& de Jong (\cite{GdJ93}), Marigo et al. (\cite{MBC98}),
Mouhcine \& Lan\c{c}on (\cite{ML02_models}).

The metallicites of the LW2000 spectra are not known. As the observed
stars are located in the solar neighbourhood, they are expected to
be representative of stars in approximately solar 
metallicity environments. The range of ages (or initial masses) in the
data is likely to correspond to a range of metallicities, as
determined by the chemical evolution of our neighbourhood.
A contamination of the sample by Population II stars is also
likely (LW2000). The calibrations of Ram\'{\i}rez et al. \cite{RSFD00}
and Frogel et al. \cite{FSRD01}, that relate [Fe/H] to the 
equivalent widths of near-IR absorption features and colours
based on Milky Way cluster red giant branch stars, would
lead to dispersed, on average slightly subsolar metallicities.
However, the calibration samples don't include LPVs, and the
observed variability of the relevant spectral signatures
in the LW2000 sample makes the results uncertain.
The LW2000 have also been compared to the synthetic
red giant spectra of Bessell et al. (\cite{BBSW89a})
and of Hauschildt et al. (\cite{HAFBA99}). Both sets
provide decent (though far from perfect) agreement with 
the observed giant star spectra and at least some of 
the LPV spectra. In the models of Hauschildt et al. (\cite{HAFBA99}),
the effect of moving from $[$M/H$]$=0
to -0.3 is so strong at low temperatures that the low
metallicity model spectra are very different from any observed 
one we have seen, in the LW2000 data or in the literature. 
With these models, one would conclude that
all the LW2000 stars have quasi-solar metallicities, but
our impression is that the effect of Z in the models is
too strong. In the model set of Bessell et al. (\cite{BBSW89a})
moving from log(Z/Z$_{\odot}$)=0 to $-1$ has
a relatively small effect on the spectra (at a given T$_{\rm eff}$,
mass and surface gravity), indicating that the observed
spectra may be appropriate over that range of metallicities.
The comparison between properties of stars in the Galactic Bulge and
in the Field supports a modest evolution of intrinsic spectral
signatures over a useful range of metallicites (Frogel et al. \cite{FWR84}).

In view of this situation, we suggest to restrict the
inclusion of metallicity effects in spectral population
synthesis models to those that are intrinsic
to stellar evolutionary tracks, and to the use of a metallicity
dependent T$_{\rm eff}$ scale for the mean O-spectra.
For instance, one may use the relations between (I-K) and
T$_{\rm eff}$ given for various metallicities by Bessell et al. \cite{BBSW91}.
The sensitivity of carbon star spectra to metallicity is small.
R.\,Gautschy (\cite{Loi01}) has computed model spectra for hydrostatic spherical
carbon stars between log(Z/Z$_{\odot}$)=0 and -0.3. The models show
changes in Z have small effects compared to changes in
the effective temperature, the C/O ratio or the $^{12}$C/$^{13}$C 
ratio. The latter are determined by the current evolutionary status
of a star, at any initial metallicity. For carbon stars,
we suggest to use the average spectra of this paper with
the same effective temperature scale at all Z. 
\medskip

\section{Summary and conclusions}
\label{concl.sec}

This paper provides convenient and representative data for stars 
on the upper asymptotic giant branch, to be used in the 
studies of cool stellar populations of galaxies.

We have used the data of LW2000 to construct sequences of average
spectra for O-rich and C-rich LPVs.
The O-rich sequence is based on (I-K), which is taken as a first 
order effective temperature indicator. Despite the large dispersion between
(I-K) (or other temperature indicators involving optical data) and near-IR
properties, a regular behaviour is observed along the average 
sequence. This would not have been the case if sorting had 
been based on near-IR indices.
For C-rich stars, both a temperature sequence and a sequence
based on the C/O ratio are presented.
For stars with thick dust envelopes (OH/IR stars and their carbon
rich equivalent), the use of reddened versions of the above
data is suggested.

In studies of stellar populations, the average spectra will be used
in connection with stellar evolution tracks or isochrones. 
The latter are expected to provide the distribution of
stars in the HR diagram, but also the onset of LPV pulsation
(i.e. when to use the average LPV spectra instead of static giant spectra),
the transition between O-rich and C-rich atmospheres, and the transition
from an optically visible object to a dust-enshrouded far-IR source.
As both the effective temperatures provided by stellar evolution
tracks for the thermally pulsing AGB and the temperatures estimated
for the average spectra are (independently) model-dependent and
uncertain, we suggest caution when using the values indicated in
this paper. The effects of reasonable changes in 
the temperature scales have been illustrated by Lan\c{c}on et al. 
(\cite{LMFS99}).
We recommend an a posteriori calibration of the relative temperature 
scales of the spectra and the evolutionary tracks, based on 
observed properties of intermediate age star clusters of known ages and 
metallicities.

Using only one parameter, the effective temperature, to characterize
LPV spectra is clearly a first order approximation (even if this
particular choice of a parameter is arguably the most appropriate one).
Pulsation period and amplitude as well as the phase in the pulsation
cycle are expected to produce systematic effects, although neither the 
LW2000 sample nor the currently available stellar models are sufficient 
to pin those down.
The period and amplitude distributions in the LW2000 data used here
are biased towards high values, when compared to the 
distribution of LPVs (including Miras and semi-regular variables)
found in star counts in the Milky Way or the Magellanic Clouds.
As only TP-AGB stars with relatively high main sequence masses will ever
contribute much to the integrated light of stellar populations, 
the problem is not as severe there. In the practice of population
synthesis calculations, the errors resulting from this
period and amplitude distribution cannot be 
disentangled from errors in the assumed onset instants of
semi-regular and Mira-type pulsation. 

\begin{acknowledgements}
We thank M.\,Scholz for providing his model spectra in digital form,
D.\,Barth\`es for pointing out a correction to standard 
overtone period-luminosity relations and providing his results in
advance of publication, F.\,Kerschbaum and B.\,Aringer for their
help with the ISO spectra they made available to us,
and P.\,Wood for his suggestions and support throughout this work. 
\end{acknowledgements}


\begin{thebibliography}{}
\bibitem[2001]{AlaMACHO01} Alard, C., the ISOGAL Collaboration, \& the
 MACHO Collaboration 2001, ApJ 552, 289
\bibitem[2000]{AlcMACHO00} Alcock, C., \& the MACHO Collaboration 2000,
 AJ 119, 2194
\bibitem[1998]{AP98} Alvarez, R., \& Plez, B., 1998, A\&A 330, 1109
\bibitem[2000]{ALPW00} Alvarez, R., Lan\c{c}on, A., Plez, B., \& Wood, P.R., 
 2000, A\&A 353, 322
\bibitem[1999]{ATO99} Aoki, W., Tsuji, T., \& Ohnaka, K. 1999,
 in IAU Symp. 191, Asymptotic
 Giant Branch Stars, Ed. Le\,Bertre, L\`ebre \& Waelkens
 (San Francisco: ASP), 175
\bibitem[1999]{AKHetal99} Aringer, B. Kerschbaum, F., Hron, J. et al. 1999
 in IAU Symp. 191, Asymptotic
 Giant Branch Stars, Ed. Le\,Bertre, L\`ebre \& Waelkens
 (San Francisco: ASP), 169
\bibitem[2001]{BL01} Barth\`es, D., \& Luri, X. 2001, A\&A, 365, 519
\bibitem[1990]{Bes90} Bessell, M.\,S. 1990, PASP 102, 1181
\bibitem[1988]{BB88} Bessell, M.\,S., \& Brett, J.\,M. 1988, PASP 100, 1134
\bibitem[1989\,a]{BBSW89a} Bessell, M.\,S., Brett, J.\,M., 
 Scholz, M., \& Wood P.\,R. 1989\,a, A\&A 213, 209
\bibitem[1989\,b]{BBSW89b} Bessell, M.\,S., Brett, J.\,M., Scholz, M., 
 \& Wood, P.\,R. 1989\,b, A\&AS 77, 1
\bibitem[1991]{BBSW91} Bessell, M.\,S., Brett, J.\,M., Scholz, M., 
 \& Wood, P.\,R. 1991, A\&AS 89, 335
\bibitem[1996]{BSW96} Bessell, M.\,S., Scholz, M., 
 \& Wood, P.\,R. 1996, A\&A 307, 481 
\bibitem[1998]{BCP98} Bessell, M.\,S., Castelli, F., \& Plez, B. 1998,
 A\&A 333, 231
\bibitem[1988]{BS88a} Boothroyd, A.\,I., \& Sackmann, I.-J 1988, ApJ 328, 632
\bibitem[1998]{BGS98} Bressan, A., Granato, G.\,L., 
 \& Silva, L. 1998, A\&A 332, 135
\bibitem[1989]{CCM89} Cardelli, J.\,A., Clayton, G.\,C., \& Mathis, J.\,S.
 1989, ApJ 345, 245
\bibitem[2000]{CvMLH00} Cioni, M.-R.\,L., van der Marel, R.\,P., 
 Loup, C., \& Habing, H.\,J. 2000, A\&A 359, 601
\bibitem[2001]{CMLetal01} Cioni, M.-R. L., Marquette, J.-B., Loup, C. et al.
 2001, A\&A 377, 945
\bibitem[1997]{Epetal97} Epchtein, N., \& the DENIS Collaboration 1997,
 The Messenger 87, 27
\bibitem[1963]{Fea63} Feast, M.\,W. 1963, MNRAS 125, 367
\bibitem[1989]{FGWC89} Feast, M.\,W., Glass, I.\,S., Whitelock, P.\,A., \&
 Catchpole, R.\,M. 1989, MNRAS 241, 375
\bibitem[1995]{FFTetal95} Ferraro, F.\,R., Fusi Pecci, F., Testa, V., et al.
 1995, MNRAS 272, 391
\bibitem[1994]{FPTetal94} Fluks, M.\,A., Plez, B., Th\'e, P.S., et al.
 1994, A\&AS 105, 311
\bibitem[1997]{FC97} Forestini, M., \& Charbonnel, C. 1997, A\&AS 123, 241
\bibitem[1982]{FW82} Fox, M.\,W., \& Wood, P.\,R. 1982, ApJ 259, 198
\bibitem[1988]{FE88} Frogel, J.\,A., \& Elias, J.\,H. 1988, ApJ 324, 823
\bibitem[1984]{FWR84} Frogel, J.\,A., Whitford, A.\,E., \& Rich, R.\,M. 1984, 
 AJ 89, 1536
\bibitem[1990]{FMB90} Frogel, J.\,A., Mould, J., \& Blanco, V.\,M. 1990, 
 ApJ 352, 96
\bibitem[2001]{FSRD01} Frogel, J.\,A., Stephens\, A.\,W., Ram\'{\i}rez\, S., 
 \& DePoy\, D.\,L. 2001, AJ, submitted (astro-ph/0101275)
\bibitem[1999]{Gau99} Gautschy, A., 1999, A\&A 349, 209
\bibitem[2001]{Loi01} Gautschy, R. (previously R. Loidl), 
 PhD Thesis, University of Vienna, Austria,
 Feb.\,2, 2001
\bibitem[1998]{GB98} Girardi, L., \& Bertelli, G. 1998, MNRAS 300, 533
\bibitem[1995]{GWCF95} Glass, E.\,S., Whitelock, P.\,A., Catchpole, R.\,M.,
 \& Feast, M.\,W. 1995, MNRAS 273, 383
\bibitem[1978]{GBSWE78} Goebel, J.\,H., Bregman, J.\,D., Strecker, D.\,W., 
 Witteborn, F.\,C., \& Erickson, E.\,F. 1978, ApJ 222, L129
\bibitem[1993]{GdJ93} Groenewegen, M.\,A.\,T., \& de Jong, T. 1993, 
 A\&A 267, 410
\bibitem[1995]{Hab95} Habing, H.\,J. 1995, Mem. Soc. Astr. It. 66, 627
\bibitem[1999]{HAFBA99} Hauschildt, P.\,H., Allard, F., Ferguson, J., 
 Baron, E., \& Alexander, D.\,R. 1999, ApJ 525, 871
\bibitem[1998]{HSW98} Hofmann, K.-H., Scholz, M., \& Wood, P.\,R. 1998, 
 A\&A 339, 846
\bibitem[1990]{HW90} Hughes, S.\,M.\,G., \& Wood, P.\,R. 1990, AJ 99,  784
\bibitem[1983]{IR83} Iben, I.\,Jr, \& Renzini, A. 1983, ARA\&A 21, 271
\bibitem[2000]{JHL00} J{\o}rgensen, U.\,G., Hron, J., \& Loidl, R. 2000, 
 A\&A 356, 253
\bibitem[2001]{JJSA01} J{\o}rgensen, U.\,G., Jensen, P., S{\o}rensen, G.\,O.,
 \& Aringer, B. 2001, A\&A 372, 249
\bibitem[1985]{GCVS85} Kholopov, P.\,N., Samus, N.\,N., Frolov, M.\,S., et al. 
 1985, General Catalogue of Variable Stars, Nauka Publishing House, Moscow
\bibitem[1986]{KH86} Kleinmann, S.\,G., \& Hall, D.\,N.\,B. 1986, ApJS 62, 501
\bibitem[1998]{Lan98} Lan\c{c}on, A. 1998, in IAU Symp. 191, Asymptotic
 Giant Branch Stars, Ed. Le\,Bertre, L\`ebre \& Waelkens 
 (San Francisco: ASP), 579
\bibitem[1992]{LRV92} Lan\c{c}on, A., \& Rocca-Volmerange, B. 1992, 
 A\&AS 96, 593
\bibitem[2000]{LM00_fluct} Lan\c{c}on, A., \& Mouhcine, M. 2000, in 
 ASP Conf. Ser. 211, Massive Stellar Clusters, ed. A. Lan\c{c}on \& C.\,M.
 Boily, 43
\bibitem[2000]{LW00} Lan\c{c}on, A., \& Wood, P.R. 2000 (LW2000), A\&AS 146, 217
\bibitem[1999]{LMFS99} Lan\c{c}on, A., Mouhcine, M., Fioc, M., \& Silva, D., 
 1999, A\&A, 344, L21
\bibitem[1999]{LSG99} Leitherer, C., Schaerer, D., Goldader, J.\,D., 
 et al. 1999, ApJS 123, 3
 A\&AS 130, 65
\bibitem[1983]{Llo83} Lloyd Evans, T. 1983, MNRAS 204, 961
\bibitem[2001]{LLJ01} Loidl, R., Lan\c{c}on, A., \& J{\o}rgensen, U.G. 2001, 
 A\&A 371, 1065 (LLJ2001)
\bibitem[1998]{Mara98} Maraston, C. 1998, MNRAS 300, 872
\bibitem[2001]{MKBBH01} Maraston, C., Kissler-Patig, M., Brodie, J.\,P.,
 Barmby, P., \& Huchra, J.P. 2001, A\&A 370, 176
\bibitem[2001]{Mar01} Marigo, P. 2001, A\&A 370, 194
\bibitem[1998]{MBC98} Marigo, P., Bressan, A., \& Chiosi, C. 1998, A\&A 331, 564
\bibitem[1999]{MGB99} Marigo, P., Girardi, L., \& Bressan, A. 
 1999, A\&A 344, 123
\bibitem[2001]{MYMO01} Matsuura, M., Yamamura, I., Murakami, H., \&
 Onaka, T. 2001, in Post-AGB Objects as a Phase of Stellar Evolution,
 ed. R.\,Szczerba \& S.\,K. G\'orny (Dordrecht: Kluwer),
 Astroph. \& Sp. Sc. Lib. 265, 433
\bibitem[2002]{ML02_models} Mouhcine, M., \& Lan\c{c}on, A. 2002\,
 A\&A, in press (Paper I)
\bibitem[2002]{MLLGS02} Mouhcine, M., Lan\c{c}on, A., Leitherer, C.,
 Groenewegen, M.\,A.\,T., \& Silva, D. 2002, A\&A, in press
\bibitem[1983]{PACFM83} Persson, S.\,E., Aaronson, M., Cohen, J.\,G., 
 Frogel, J.\,A., \& Matthews, K. 1983, ApJ 266, 105
\bibitem[1998]{Pick98} Pickles, A.\,J. 1998, PASP 110, 863
\bibitem[2000]{RSFD00} Ram\'{\i}rez, S.\,V., Stephens, A.\,W., Frogel, J.\,A., 
 \& DePoy, D.\,L. 2000, AJ 120, 833
\bibitem[1992]{Ren92} Renzini, A. 1992, in IAU Symp. 149, The Stellar
 Populations of Galaxies, ed. B. Barbuy \& A. Renzini (Dordrecht: Kluwer), 325
\bibitem[1981]{RV81} Renzini, A., \& Voli, M. 1981, A\&A 94, 175
\bibitem[1986]{RB86} Renzini, A., \& Buzzoni, A. 1986, in Spectral Evolution
 of Galaxies, ed. C. Chiosi \& A. Renzini (Dordrecht: Reidel), 195
 galaxies
\bibitem[1987]{ST87} Scholz, M., \& Takeda, Y. 1987, A\&A 186, 200 (erratum
 1988, A\&A 196, 342)
\bibitem[1969]{SW69} Spinrad, H., \& Wing, R.F. 1969, ARA\&A 7, 249
\bibitem[1991]{TFW91} Terndrup, D.\,M., Frogel, J.\,A., \& Whitford, A.\,E. 
 1991, ApJ 378, 742
\bibitem[1997]{vBetal97} van Belle, G.\,T., Dyck, H.\,M., Thompson, R.\,R.,
 Benson, J.\,A., \& Kannappan, S.\,J. 1997, AJ 114, 2150
\bibitem[1993]{VW93} Vassiliadis, E., \& Wood, P.\,R. 1993, ApJ 413, 641
\bibitem[1998]{WG98} Wagenhuber, J., \& Groenewegen, M.\,A.\,T. 1998, 
 A\&A 340, 183
\bibitem[1996]{WT96} Wagenhuber, J., \& Tuchman, Y. 1996, A\&A 311, 509
\bibitem[2001]{WN01} Weinberg, M.\,D., \& Nikolaev, S. 2001, ApJ 548, 712
\bibitem[1990]{Wo90} Wood, P.R. 1990, in From Miras to Planetary Nebulae:
 Which Path for Stellar Evolution?, ed. M.-O. Mennessier \& A. Omont
 (Gif sur Yvette: Editions Fronti\`eres), 67
\bibitem[2000]{Wo00} Wood, P.R. 2000, PASA 17, 18
\bibitem[1999]{WMACHO99} Wood P.,R., and the MACHO Collaboration 1999, in
 IAU Symp. 191, Asymptotic Giant Branch Stars, ed. T. Le Bertre, A. L\`{e}bre 
 \& C. Waelkens (San Francisco: ASP), 151
\bibitem[1998]{XDC98} Xiong, D.\,R., Deng, L., \& Cheng, Q.\,L. 1998, 
 ApJ 499, 355
\end{thebibliography}
\end{document}